# Accelerating Discovery of Ternary Chiral Materials via Large-Scale Random Crystal Structure Prediction

Jiexi Song [a, #], Diwei Shi [b, #], Fengyuan Xuan [a,*] and Chongde Cao [c,*]

a Suzhou Laboratory, Suzhou 215123, China
b School of Naval Architecture and Maritime, Zhejiang Ocean University, Zhoushan 316022, China
c School of Physical Science and Technology, Northwestern Polytechnical University, Xian 710072, China

#: Jiexi Song and Diwei Shi contributed equally

Author to whom correspondence should be addressed:
xuanfy@szlab.ac.cn; caocd@nwpu.edu.cn

## Abstract

Chiral inorganic crystals, particularly semiconductors with Weyl points near the band edges or semimetals hosting Weyl points at the Fermi level, have attracted considerable interest, yet they remain scarce in existing materials databases. This study presents a prediction pathway by combining universal machine learning interatomic potentials (uMLIPs) for high-throughput structure optimization with the broad exploration capability of random structure search (RSS), enabling large-scale crystal structure prediction in ternary systems with variable compositions, followed by targeted screening for chiral space groups. Through uMLIP-based high-throughput optimization and stability assessment, a large number of potentially stable phases were identified from over 20 million randomly generated chiral structures. First-principles validation further confirmed more than 260 chiral inorganic crystals with potential applications in topological properties, nonlinear optics, and superconductivity. Some of these materials exhibit notable quantum phenomena, such as the nonlinear Hall effect driven by Berry curvature dipole, quantum metric and symmetry-protected sixfold degenerate topological points, long Fermi arcs, and large magnetoresistance. This work substantially expands the pool of candidate chiral functional materials and offers a scalable strategy for predicting ternary material systems.

## Introduction

Chiral crystals, characterized by the absence of mirror planes, inversion centers, and improper rotation axes, underpin the manifestation of unusual quasiparticles, including Kramers-Weyl and multifold-degenerate fermions, [1-4] as well as bosonic excitations such as excitons, phonons, and magnons with chiral character.[5-6] Notably, for nonmagnetic chiral crystals, their electronic states demonstrate substantial promise for exploiting fascinating quantum phenomena. For instance, recently discovered chiral semimetals systems like CoSi exhibit long Fermi arcs across nearly the whole Brillouin-zone.[7-8] These chiral topological materials can also host robust Weyl fermions protected by time-reversal symmetry, and demonstrate phenomena such as the gyrotropic magnetic effect.[9] Another key advantage of chiral topological materials is their potential for nonlinear optical (NLO) devices, stemming from the noncentrosymmetric nature inherent to chiral structures. Specifically, chiral topological materials exhibit quantum geometry which leads to quantized circular photogalvanic effect (CPGE) and a giant bulk photovoltaic effect (BPVE) extending from terahertz (THz) to the visible spectrum.[10-11]

Despite these remarkable quantum phenomena and NLO responses, experimentally confirmed chiral inorganic candidates remain exceptionally scarce. Notably, a high-throughput screening of over 210,000 synthesized compounds in the ICSD database identified specifically 46 viable nonmagnetic Weyl semimetals with promising topological properties and nearly clean Fermi surfaces, of which only nine compounds possess chiral symmetry.[11] Moreover, a recent high-throughput research demonstrates that even when applying a more relaxed stable criterion (0.5 eV/atom vs. the typical 0.2 eV/atom) and focusing on specific high-fold degeneracy types, scanning the entire Materials Project database (>140,000 entries) identifies only 146 chiral topological semimetals.[12] These high-throughput screenings, while successfully identifying a handful of high-quality chiral Weyl semimetals, underscore the scarcity of such ideal candidates. Given that the vast majority of Weyl semimetals lack these advantageous characteristics, this scarcity emphasizes the pressing need for systematic discovery approaches. A promising path forward lies in expanding the search into the space of unknown crystal structures, which is a main focus of modern computational materials discovery. On the other hand, searching for functional materials in the space of unknown crystal structures is the main focus of computational materials discovery studies. Random Structure Search (RSS) is a non-iterative method instrumental in discovering numerous distinctive crystals, such as the ambient-pressure high-*T*c $Mg_2IrH_7$[13] and high-*T*c Perovskite Hydride below 10 GPa.[14] Yet, as the target systems expand to encompass a continuous range of stoichiometries and dramatically increasing numbers of elements, potentially spanning significant portions of the periodic table, the prohibitive computational cost of DFT critically limits the application of RSS. Fortunately, rapidly developing universal machine learning potentials (uMLIP), delivering superior efficiency and sufficient accuracy, pave the way for large-scale random crystal structure prediction.[15-18]

In this work, we performed a high-throughput crystal structure prediction research for three different configurations $M_{1\text{-}3}X_{1\text{-}3}Hal_{1\text{-}3}$, $M_{1\text{-}3}M'_{1\text{-}3}B_{1\text{-}3}$ and $TM_{1\text{-}4}M_{1\text{-}4}Ch_{1\text{-}4}$ with variable elements covering most of the elements in the periodic table. Here, M, M' and TM denote metal elements, X represents some nonmetals (N, P, O, S, Se), Hal stands for halogens (F, Cl, Br, I), B is boron and Ch denotes chalcogens (S, Se, Te). Employing a strict strategy based on Wyckoff site occupancy and constrained by symmetry, approximately 20 million crystal structures were randomly generated across 65 Sohncke space groups. The uMLIP was then employed to conduct large-scale structural optimization of these 20 million crystal structures. Following rigorous stability and density functional theory (DFT) evaluations, over 260 dynamically stable nonmagnetic chiral materials were identified, among which more than 60 also exhibit promising thermodynamic stability ($E_{hull} \leq 0.05$ eV/atom), and related NLO and superconducting properties were systematically explored. This work integrates large-scale crystal structure prediction with the rapidity and accuracy of uMLIP, discovering numerous potential chiral topological materials. The developed methodology is readily extendable to crystal structure prediction in other multinary variable-composition systems.

# Results

## Screening workflow

In this work, we define chiral crystals as those belonging to the 65 Sohncke space groups, which are noncentrosymmetric and contain no mirror or improper rotation symmetries.. The screening workflow is shown in Fig. 1. Our workflow can be summarized into three primary modules. The initial module encompasses crystal structure generation. Within this module, all starting structures are rigorously constrained by their space group number to generate configurations satisfying Wyckoff site occupancy requirements. Generation was restricted to the 65 Sohncke space groups, explicitly excluding the *P*1 symmetry. Consequently, the initial space group list comprised numbers 3-5, 16-24, 75-80, 89-98, 143-146, 149-155, 168-173, 177-182, 195-199, and 207-214 (totaling 64 groups). For each distinct composition (e.g., $Pd_1N_1F_1$), the algorithm randomly generates 30 structures within each of the aforementioned 64 space groups. Additional constraints were imposed during structure generation, the number of atoms was capped at 24 per unit cell. To enhance structural diversity for a given composition, a multiplicity parameter was introduced. For instance, in an equiatomic system like PdNF, structures were generated across varying atom counts with the same stoichiometry ($(PdNF)_x$, where x ranges from 1 to 8). This functionality was implemented via an extension of the PyXtal package.[19] Due to the stringent combinatorial restrictions imposed by Wyckoff site assignments in ternary systems, structure generation within certain space groups under specific conditions (multiplicity plus atom count) could fail. In such instances, if repeated attempts (across all valid multiplicity plus atom count combinations) within a given space group proved unsuccessful, that space group was skipped. Employing this algorithmic strategy, we conducted large-scale, symmetry-constrained random crystal structure generation across three distinct material systems, yielding a total of nearly 20 million structures.

The second core module of our screening algorithm implements large-scale crystal structure optimization and stability assessment. For the generated structures, we performed high-throughput crystal structure optimization using the mattersim-v1.0.0-1M foundation model,[20] leveraging its favorable accuracy and computational efficiency. Optimization proceeded until atomic forces converged below 0.05 eV/Å. Subsequently, the lowest-energy structure for each composition was screened out. Energy above hull ($E_{hull}$) were then calculated for these structures using reference data from the Materials Project (MP) database.[21] A deliberately conservative $E_{hull}$ threshold of 0.3 eV/atom was applied at this stage. This initial screening yielded over 4700 distinct compositions/element sets exhibiting predicted stability. These candidate structures underwent symmetrization analysis, resulting in the exclusion of any system adopting a centrosymmetric configuration post-optimization. Recognizing that thermodynamic stability (fixed on $E_{hull}$) alone is insufficient and may retain pseudo-stable configurations, phonon dispersion calculations are essential for assessing dynamic stability in periodic crystals. Given the prohibitive computational cost of phonon calculations at this scale using ab initio methods, we computed phonon spectra for all remaining structures using uMLIP, systematically eliminating those exhibiting imaginary frequencies. Benchmarking against density functional theory (DFT) with the PBE functional confirms uMLIP's reliability for qualitative phonon stability assessment.[22] Structures satisfying both thermodynamic and dynamic criteria were then subjected to homologous substitution. Within our algorithm, if a composition (e.g., PdNF) passed screening, we generated derivative structures by substituting Pd with (Ni, Pd, Pt), N with (N, P, As, Sb, Bi), and F with (F, Cl, Br, I). This expanded set underwent the full screening cascade, re-optimization (force convergence ≤ 0.01 eV/Å), $E_{hull}$ calculation (with 0.2 eV/atom threshold), and phonon stability screening. All structures emerging from this substitution protocol were subsequently re-optimized using DFT for final validation. This step proved critical, as exemplified in Fig. 2c, where DFT re-optimization increased the convex hull energy above the 0.2 eV/atom threshold for some structures deemed stable by uMLIP alone. Following DFT validation, approximately 300 stable compounds were confirmed. Finally, symmetry analysis identified over 260 of these as chiral crystals.

Figs 2.b visualizes the t-SNE[23] projection maps derived from the Smooth Overlap of Atomic Positions (SOAP) descriptors[24] for a randomly selected set of 20,000 generated initial structures alongside their optimized counterparts using uMLIP. This visualization emphasizes the critical necessity of structural optimization, while direct energy prediction from initial structures via uMLIP to identify low-energy

configurations is undesirable, as many structures undergo a change in space group symmetry during optimization. Furthermore, the substantial spatial overlap observed between the pre- and post-optimization feature distributions suggests that the symmetry-constrained crystal structure generation approach employed is well-suited for predicting chiral materials within specific space groups. Following our predefined target distribution of initial material systems, we predicted key electronic and functional properties for the aforementioned screened chiral inorganic crystals. These predictions included NLO response and superconducting properties, tailored to the specific material class. For the ternary boride systems under investigation, the primary computational focus encompassed the superconducting critical temperature (*Tc*) and Vickers hardness (*Hv*). In addition, for ternary semiconductors incorporating halogen elements, the emphasis shifted to calculating their NLO response coefficients.

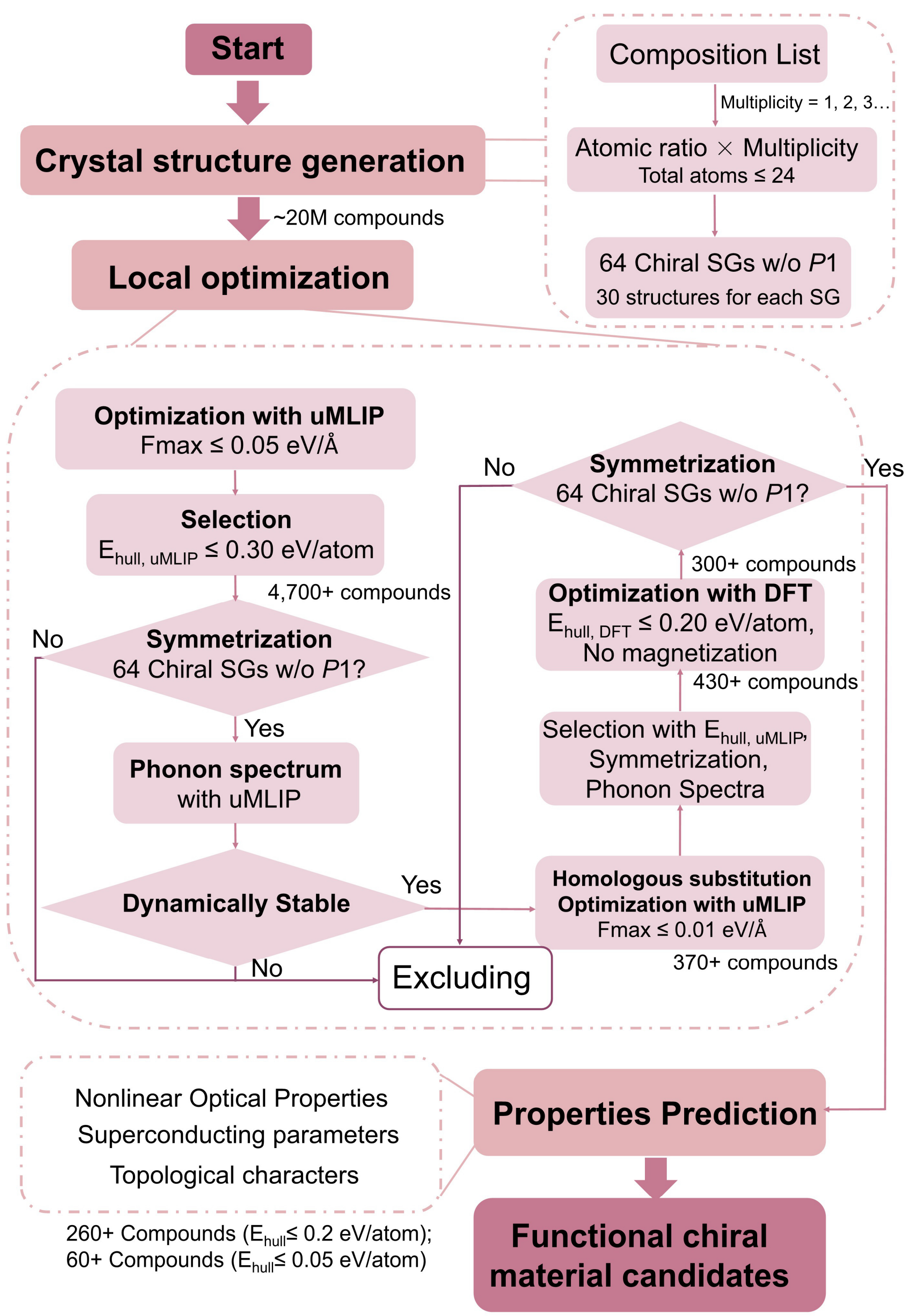

**Fig. 1 Workflow for the symmetry constrained large scale crystal structure prediction**. Chiral crystal structures were initially generated via stochastic sampling and optimized using a universal machine learning interatomic potential (uMLIP). The resulting optimal structures underwent further refinement via density functional theory (DFT). Subsequent screening based on energy above the convex hull ($E_{hull}$), phonon stability (absence of imaginary frequencies), and symmetry analysis yielded over 260 chiral materials. These identified compounds were systematically characterized to assess their topological properties, nonlinear optical response, and superconducting behavior.

## The distribution of predicted chiral compounds

Figs 2.a portrays the distribution of the predicted chiral structures satisfying the stability criteria, alongside representative crystal structure images. It is noteworthy that our workflow also yielded other stable and structurally intriguing materials outside the chiral space group scope, a prominent example, *Pnnm*-PdNF, is also displayed in Fig. 2a for its novelty woven from one-dimensional (1D) chains despite being non-chiral. The results reveal that the space group $P2_12_12_1$ contains the highest number of structures, followed by $P2_1$, $P3_1$ & $P3_2$, $P4_1$ & $P4_3$, and $P2_13$. It should be pointed out that this work investigated Sohncke space groups. Strictly enantiomorphic space groups, possessing distinct left-handed and right-handed counterparts, are exemplified by types like $P4_1$ (where the $4_1$ and $4_3$ screw axes form an enantiomorphic pair). Only 22 of the 230 space groups satisfy this strict enantiomorphism requirement.[25] Space groups containing screw axes such as $2_1$ or $4_2$ are self-enantiomorphic, meaning they do not possess distinct enantiomorphs. Nevertheless, these non-strictly enantiomorphic chiral space groups remain crucial candidates for exploring phenomena like Kramers-Weyl fermions and high-fold degenerate fermions.[12, 25]

Representative crystal structures are displayed. Boride structures generally exhibit denser atomic packing compared to others. Halogen-containing systems, such as $P3_12_1$-$OsI_3O$, display a quasi-two-dimensional architecture, while structures like $P4_1$-$ReNF_3$, $P4_22_2$-$IrN_2F_2$, and $P3_1$-$CuN_2Br$ contain significant holes. As mentioned, the quasi-2D bulk structure *Pnnm*-PdNF, woven from 1D atomic chains, is a non-chiral but dynamically stable compound identified during our screening. Phonon dispersion calculations confirmed dynamic stability for both the full 3D bulk structure and the isolated 1D atomic chain (see Figs. S2). The structural evolution and enhanced stability from the 1D chain to the 3D bulk indicate that the thermodynamic properties of this material merit further investigation. Figs 2.c further demonstrates that structures identified within the $M_xX_yHal_z$ system generally exhibit superior thermodynamic stability, with a median formation energy above the convex hull below 0.1 eV/atom and contains the largest number of structures below 0.2 eV/atom. While the $M_xM'_yB_z$ system contains the second number of structures below 0.2 eV/atom, many around 0.1 eV/atom. In contrast, a significant portion of the $TM_xM_yCh_z$ structures became thermodynamically unstable after DFT relaxation. Notably, the $TM_xM_yCh_z$ system yielded only 7 structures below 0.1 eV/atom, substantially fewer than the other two compositional categories explored.

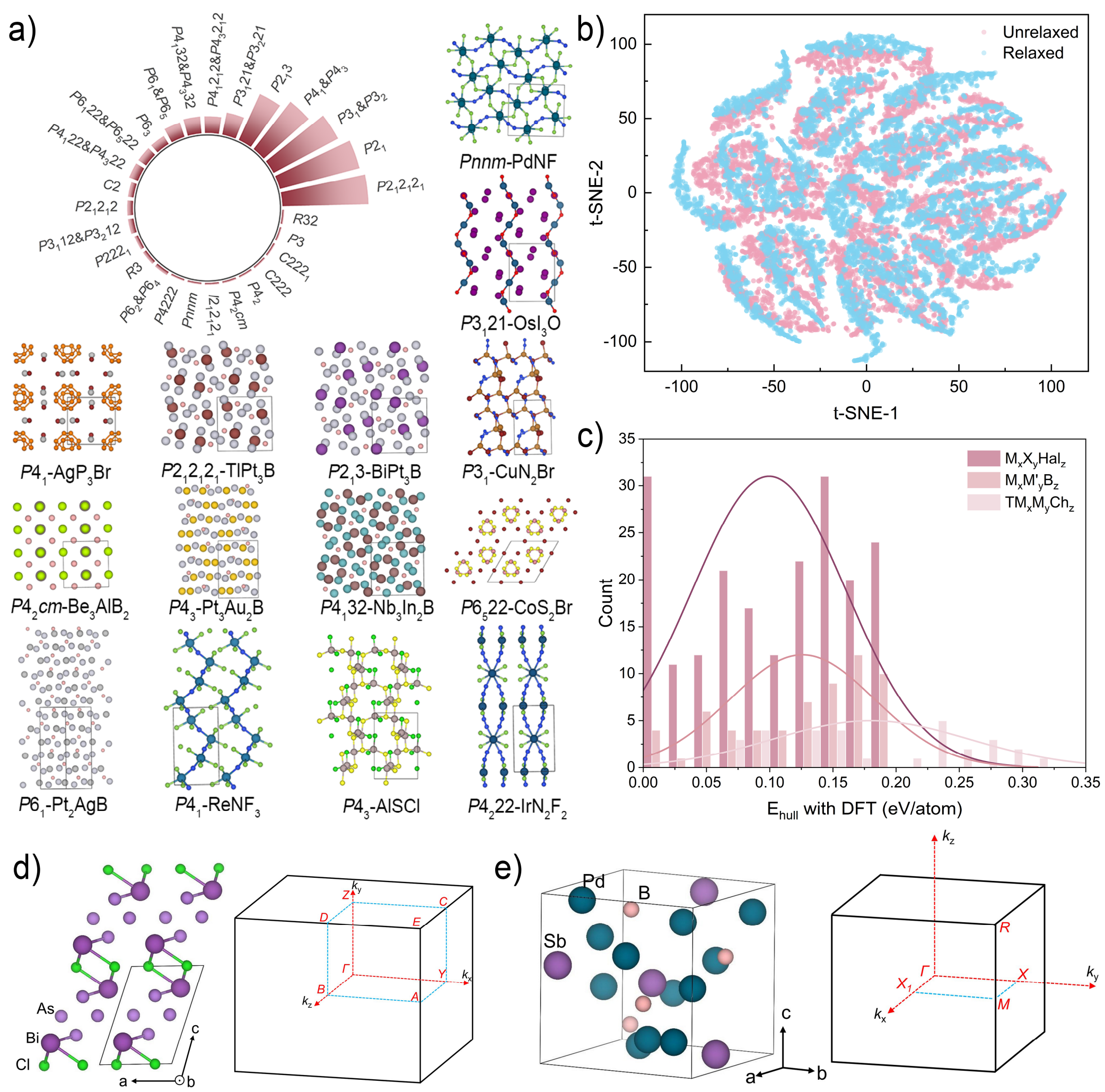

**Fig. 2 Overview of screening results.** a) Statistical distribution of predicted structures across Sohncke space groups, the prototype labeled '*Pnnm*-PdNF' is a non-chiral but stable compound discovered incidentally. Representative structural prototypes are displayed. b) t-SNE visualization of structural diversity: Comparison of 20,000 randomly sampled configurations in initial and uMLIP-optimized states using the Smooth Overlap of Atomic Positions (SOAP) descriptors.[24] c) Population statistics for structures pre-screened by uMLIP and subsequently validated via DFT re-optimization and self-consistent calculations. The structure configuration and Brillouin zone of d) $P2_1$-$BiAs_2Cl$ and e) $P2_13$-$Pd_3SbB$.

The elemental distributions for the three systems explored in this study are presented in Figs. 3.a, Figs. 4.a, and Figs. S3.a, respectively. The rationale for selecting the element types can be summarized as follows: Firstly, we investigated the $M_xX_yHal_z$ system, inspired by common NLO crystals containing halogens.[26-27] In this system, compounds consist of one metal element and two distinct non-metal elements. Element

M encompasses the majority of elements from the periodic table, position Hal is fixed as a halogen element, and X was selected from five non-metal elements. Our primary objectives within this system were to identify potential NLO semiconductors and systems exhibiting Kramers-Weyl fermions near the band edges. Secondly, motivated by the prevalence of borides under ambient conditions, we designed the $M_xM'_yB_z$ system. This system comprises compounds formed by two metal elements and boron. The principal aim was to discover potential superconductors and multiply degenerate topological semimetals within this family. Additionally, we explored select ternary chalcogenides, specifically the $TM_xM_yCh_z$ system.

Following large-scale structural optimization of nearly 20 million structures in total, we extracted results pertaining to the lowest-energy optimized structure for each composition. The $E_{hull}$ for these structures was subsequently computed combined with the MP database and the details can be found in our supplementary materials. The statistical distributions of $E_{hull}$ for the three systems are presented in Figs. 3.b, Figs. 4.b, and Figs. S3.b, respectively. As evident from Figs. 3.b, the $E_{hull}$ values for the $M_xX_yHal_z$ system are predominantly clustered around 0.2 eV/atom, with only a very small fraction of structures exhibiting values over 0.5 eV/atom. The distribution shifts notably for the $M_xM'_yB_z$ boride system, where the proportion of structures with $E_{hull}$ around 0.3 eV/atom increases significantly. Furthermore, among this subset of thermodynamically metastable structures, their $E_{hull}$ values are primarily distributed within the 0.1–0.2 eV/atom range. Regarding the explored $TM_xM_yCh_z$ system, we identified structures with $E_{hull}$ values only between 0.1 and 0.2 eV/atom, indicating no thermodynamically stable phases ($E_{hull}$ = 0) were found within this system. This absence may be attributed to the narrower elemental scope in the $TM_xM_yCh_z$ system or an insufficient number of initial structures in our calculations. Furthermore, the electronic structures of the predicted $TM_xM_yCh_z$ structures did not exhibit particularly noteworthy properties. Consequently, our subsequent property calculations prioritized the exploration of the $M_xX_yHal_z$ and $M_xM'_yB_z$ systems. We further compared the crystal structures between our predicted compounds and the database, and the relevant comparisons can be found in the supplementary materials, see Fig S1, Table S1 and Table S3. And the energy difference between DFT energy and uMLIP energy for predicted stable chiral structures in this work can be found in Fig.S4, the RMSE of energy is 0.047 eV/atom.

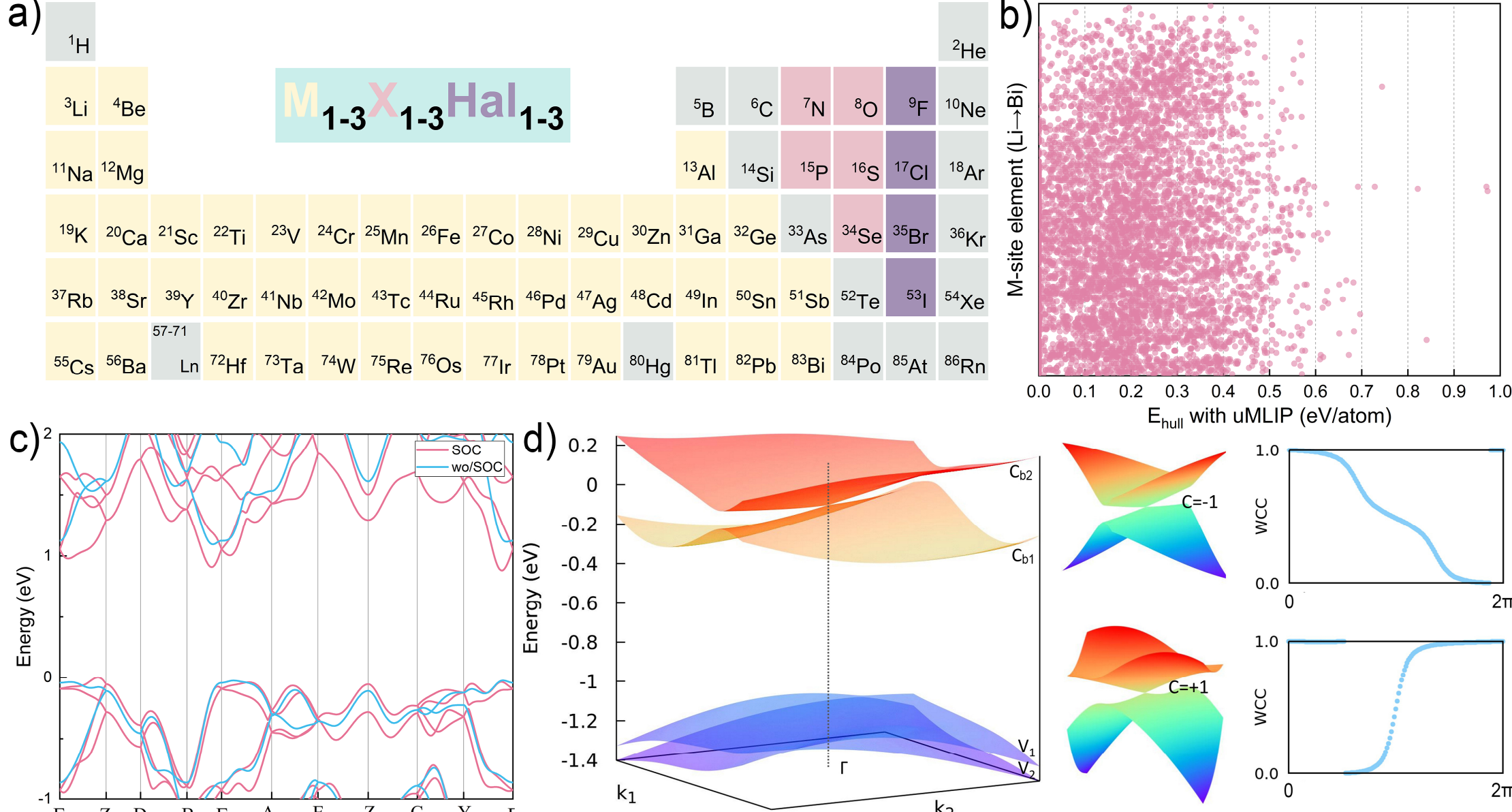

**Fig. 3 Screening results of $M_xX_yHal_z$ system.** a) Elemental selection for the $M_xX_yHal_z$ system. b) Distribution of energies above convex hull ($E_{hull}$) for the lowest-energy structure extracted for each composition within the $M_xX_yHal_z$ system. c) Band structure of $P2_1$-$BiAs_2Cl$ with and without spin-orbit coupling (SOC) considered. d) Three-dimensional (3D) band structure near the Γ-point showing conduction and valence band-edge crossings for $P2_1$-$BiAs_2Cl$, alongside plots of the topological charge and Wilson loop associated with these crossings.

Ultimately, the information for compounds within the $M_xX_yHal_z$ system exhibiting

$E_{hull}$ below 0.2 eV/atom, as determined by DFT calculations, are listed in Table S1. Among these, we predict 23 thermodynamically stable phases ($E_{hull}$ = 0) compared to existing structures in Materials Project database, and we further check their stability compared to structures existing in OQMD database. Our analysis revealed that many compounds, for example, $P2_12_12_1$-$AlSe_2Cl$, $P2_1$-$ZrSe_3Br_2$, $P3_1$-$InS_3I_3$ $P6_522$-$RhS_2Cl$ and $P6_522$-$CoS_2Br$ were absent from existing known databases. Moreover, our analysis identifies several configurations with structural prototypes already present in the OQMD database. While some of these known structures, such as the one for $ZrS_2Br_2$, possess lower energy, a substantial fraction of our predictions are energetically more favorable. For instance, the predicted $P2_13$-$AgBrO_3$ structure exhibits a lower energy (~ -0.03 eV/atom) than its known $R3c$ counterpart in the OQMD. This result underscores the promise of high-throughput, variable-element random structure prediction in identifying previously unexplored compounds with stable configurations. Among the predicted structures, many compounds exhibit semiconducting behavior, while the remainder are metallic. We then computed the second-harmonic generation (SHG) coefficients $\chi^{(2)}_{\alpha\beta\gamma}$, for the promising candidates with an $E_{hull} \leq 20$ meV/atom (see Table 1 and Table S2). Results indicate notably large SHG coefficients for $P2_1$-$BiAs_2Cl$ and $P3_1$-$InS_3I_3$, with values of 339.16 pm/V and 235.54 pm/V, respectively. Furthermore, we notably identified some structures harboring Kramers-Weyl fermions near the band edges, which is a rare phenomenon previously reported only in chiral elemental Tellurium and $ZrGeTe_4$.[28-29] This specific structure will be discussed in detail later. Additionally, we investigated the electronic structure of other metallic structures within $E_{hull} = 0$. This exploration revealed an np-type "cold metal",[30-31] $P4_1$-$ReNF_3$, characterized by the presence of band gaps both above and below the Fermi level, as presented in Figs. S5.a-b. Given this unusual electronic configuration, the material's properties could be leveraged for applications in areas such as electronic transport and thermoelectric conversion.[31]

Relevant information for structures within the $M_xM'_yB_z$ system possessing an $E_{hull}$ below 0.2 eV/atom, as calculated using DFT, can be found in Table S3. For the $M_xM'_yB_z$ system, after verifying the non-magnetic nature of these structures, we calculated their superconducting transition temperature (*Tc*) and Vickers hardness (*Hv*). We observed significant variation in the calculated hardness values of the ternary borides. For instance, the structure predicted to have the highest *Hv* is $P2_12_12_1$-$Be_3CuB$, reaching 40.26 GPa, approaching the hardness of typical superhard borides such as $CrB_4$ (48 GPa)[32] and $WB_4$ (46.3 GPa).[33] In contrast, some Au-containing structures, such as $P4_3$-$Pt_3Au_2B$ and $P4_132$-$Li_2Au_3B$, exhibit much lower *Hv* of only 0.476 GPa and 0.515 GPa, respectively. The *Tc* also displayed considerable disparity. The majority of structures are non-superconducting. The three structures with the highest predicted *Tc* values are $P4_3$-$Be_3AlB$ (15.93 K), $P4_1$-$Li_3Al_2B$ (9.68 K), and $P2_1$-$Tl_3Pt_2B$ (5.76 K). The logarithmic plot of *Tc* versus *Hv* for structures exhibiting non-zero *Tc* is presented in Figure 4c. This figure reveals that among the predicted boride superconductors, $P4_3$-$Be_3AlB$ possesses both the highest superconducting transition temperature and a relatively high *Hv* (21.62 GPa). This combination suggests its potential as a superhard superconducting material.

**Table 1**. The second-harmonic generation (SHG) coefficients $\chi^{(2)}_{\alpha\beta\gamma}$ (pm/V) of semiconductors in $M_xX_yHal_z$ system with relatively high nonlinear optical responses and $E_{hull} \leq 20$ meV/atom.

| $\chi^{(2)}_{\alpha\beta\gamma}$ | $P2_1$-$BiAs_2Cl$ | $P3_1$-$InS_3I_3$ | $P2_1$-$ZrSe_3Br_2$ | $P6_522$-$RhS_2Cl$ | $P6_522$-$CoS_2Br$ | $P3_1$-$InS_3Br_3$ | $P2_12_12_1$-$TiSe_2I_2$ | $P2_1$-$TiSe_3Br_2$ |
|---|---|---|---|---|---|---|---|---|
| 111 | | -25.43 | | | | -4.11 | | |
| 112 | -49.18 | -17.06 | -207.10 | | | -8.00 | | -161.83 |
| 113 | | -13.72 | | | | -8.53 | | |
| 123 | 99.02 | 1.52 | 146.78 | -201.43 | -201.14 | 15.08 | -166.55 | 107.88 |
| 211 | -256.04 | | 3.01 | | | | | 63.32 |
| 213 | -339.16 | | 48.83 | | | | 27.15 | 18.85 |
| 222 | 184.4 | | 88.64 | | | | | 103.20 |
| 233 | -146.44 | | 58.11 | | | | | 53.28 |
| 311 | | 59.04 | | | | 35.61 | | |
| 312 | -56.17 | | 21.43 | | | | -73.98 | -80.39 |
| 323 | 152.98 | | 132.23 | | | | | 158.80 |
| 333 | | -235.54 | | | | -190.79 | | |

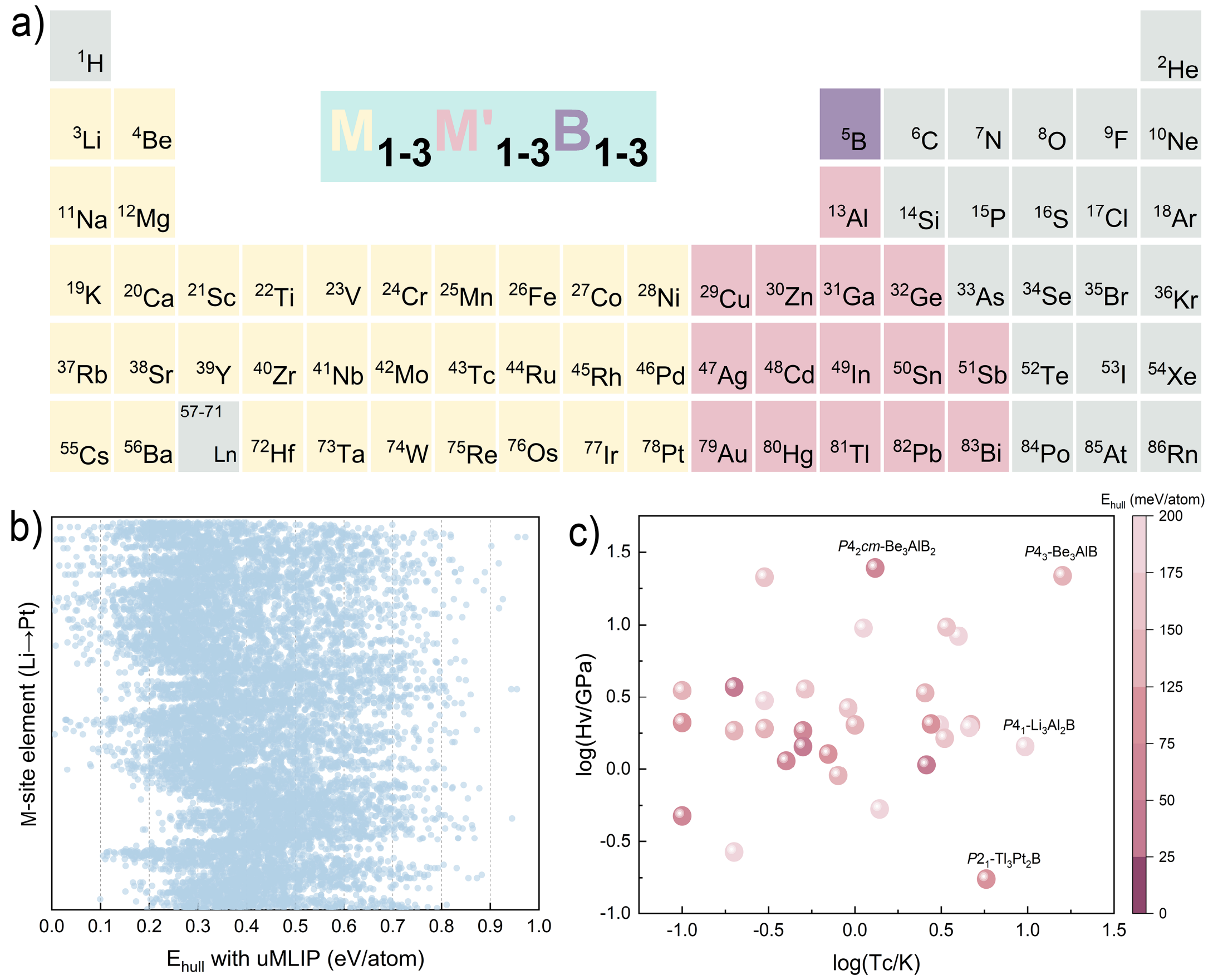

**Fig. 4 Screening results of $M_xM'_yB_z$ system.** a) Elemental selection for the $M_xM'_yB_z$ system. b) Distribution of $E_{hull}$ for the lowest-energy structure extracted for each composition within the $M_xM'_yB_z$ system. c) A log-statistical graph of superconducting-Vickers hardness for structures with non-zero superconducting transition temperatures in the $M_xM'_yB_z$ system.

## Representative materials

### A. *P*$2_1$-$BiAs_2Cl$: Candidate for Intrinsic Nonlinear Hall Effect Driven by Berry Curvature Dipole and Quantum Metric

The structure configuration and Brillouin zone of *P*$2_1$-$BiAs_2Cl$ was displayed in Figs 2d. As theoretical studies indicate, a particularly fascinating characteristic of non-magnetic chiral crystals lies in their potential to host Weyl points protected by time-reversal symmetry at high-symmetry points in momentum space when spin-orbit coupling is considered.[4] Unlike conventional Weyl points, which typically emerge accidentally due to magnetic order or crystal symmetries breaking, these crossings, termed Kramers-Weyl fermions, exhibit inherent robustness.[4] This stability arises without requiring external magnetic fields or doping, significantly enhancing their suitability for quantum device applications. Moreover, due to the presence of Berry curvature dipole (BCD),[34-38] Kramers-Weyl systems may exhibit an intrinsic second-order nonlinear Hall effect. Their inherent rectification properties hold significant promise for applications such as energy harvesting and wireless charging and THz sensing.[39] In contrast, other systems like $WTe_2$ require the application of large external electric fields to lift electronic degeneracy,[37-38] suffering from limited tunability which restricts their application potential Furthermore, to date, chiral semiconducting materials featuring Kramers-Weyl points at band edges have only been reported in Tellurium and $ZrGeTe_4$, with nonlinear Hall effects observed solely in Tellurium.[28-29] This scarcity motivates our systematic exploration of the band structures of all predicted chiral semiconductors within the $M_xX_yHal_z$ family.

Notably, we identified a compound, $P2_1$-$BiAs_2Cl$, within these systems exhibiting a compound featuring obvious Kramers-Weyl points at the band edges near the Γ point. As shown in Figs 3c, the conduction band near Γ displays a valley-like dispersion without SOC, and both the conduction and valence bands are remarkably "clean", providing an excellent platform for carrier manipulation. Upon including SOC, the bands at Γ undergo degeneracy lifting, resulting in a two-fold degenerate Kramers-Weyl point. Degeneracy lifting also occurs near the ***A*** point. However, the Kramers-Weyl point at ***A*** in $P2_1$-$BiAs_2Cl$ is located at a higher/deeper energy compared to the one at Γ. Consequently, as depicted in the upper panel of Figs S3.c, when the Fermi level is tuned to slightly intersect the conduction band, the underlying transport properties of $P2_1$-$BiAs_2Cl$ are dominated by the bands near Γ. This pristine electronic structure undoubtedly establishes $P2_1$-$BiAs_2Cl$ as a prime experimental candidate for verification. For comparison, we present in Figs S5.c-d the band structure of $P2_1$-$BiP_2Br$, a predicted isostructural compound with elemental substitution. In contrast to $P2_1$-$BiAs_2Cl$, the band weight near the A point in $P2_1$-$BiP_2Br$ is comparable to that near Γ, resulting in a "dirty" band structure.

Furthermore, Figs 3c reveals significant anisotropy in the band curvature around the Γ-point Kramers-Weyl point in $P2_1$-$BiAs_2Cl$, a feature also evident in the 3D band structure shown in Figs 3d. We further computed the topological charges associated with these two band-edge, two-fold degenerate crossing points, confirming their topological nature. Additionally, spin-projected bands were also calculated around Γ for $P2_1$-$BiAs_2Cl$, verifying the characteristic spin-momentum locking, these results can be found in Figs S6.d-i.

We then conducted a detailed investigation of the nonlinear response in $P2_1$-$BiAs_2Cl$. Initially, we constructed maximally localized Wannier functions (fitting results are presented in Figs S3.d) and computed its shift current tensor using a high-density k-point grid. The two largest components of the shift current tensor are shown in Figs 5a. Notably, $\sigma_{yxx}$ for $P2_1$-$BiAs_2Cl$ exhibits a pronounced peak in the visible spectrum, with a magnitude reaching 30 μA/V², indicating that this compound is a promising nonlinear optical material.

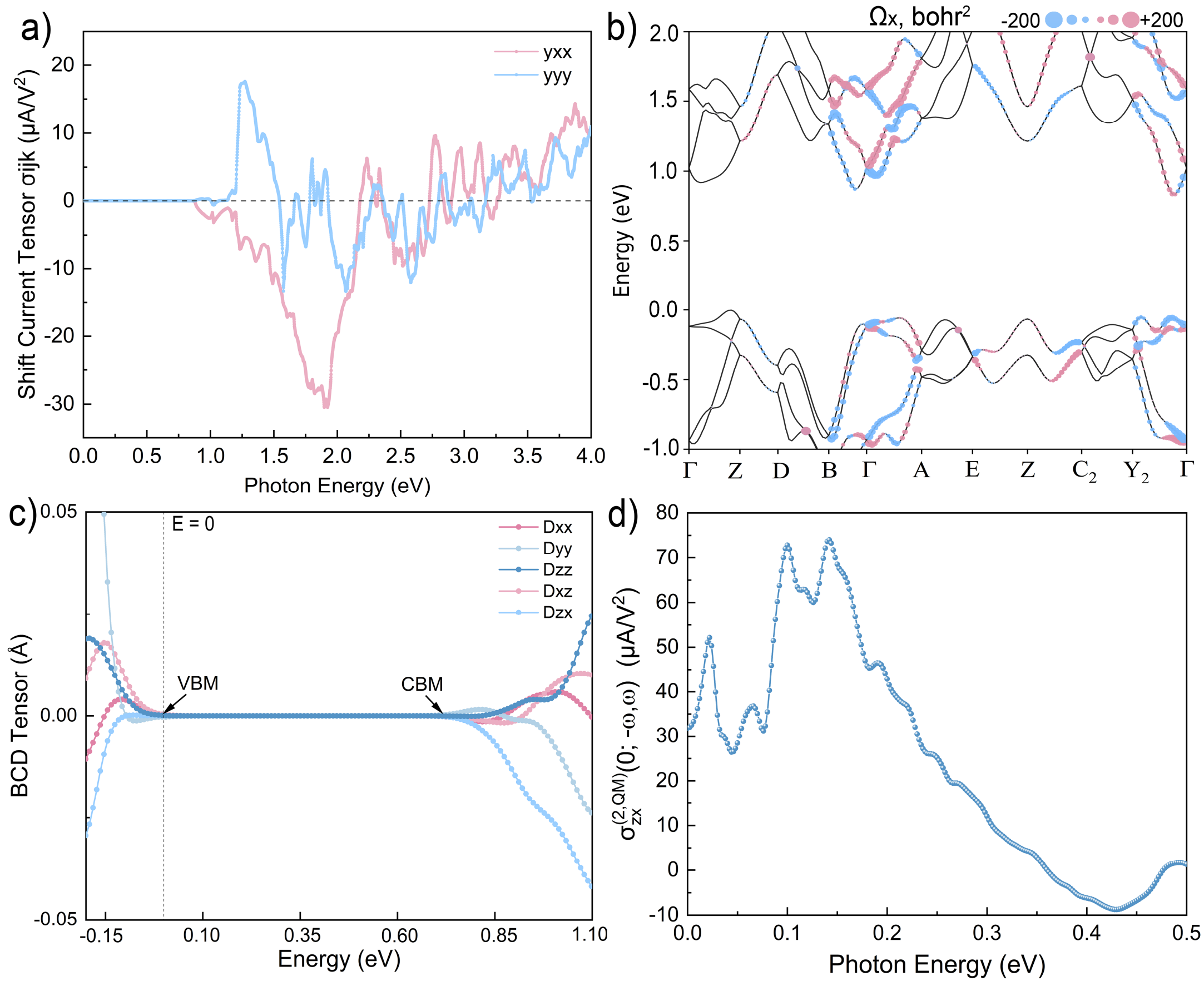

**Fig. 5 Nonlinear properties of $P2_1$-$BiAs_2Cl$.** a) Shift current tensor, b) Band structure with Berry curvature $\Omega_x$ projection, c) Non-zero Berry curvature dipole (BCD) tensor components and d) Nonlinear Hall current from quantum metric (QM) contribution.

Subsequently, we calculated band-resolved Berry curvature components $\Omega_i$ (i = x, y, z) for $P2_1$-$BiAs_2Cl$, the results are presented in Figs 5b, the lower panel of Figure S2.c, and Figs S6. a-c. Taking $\Omega_x$ as an example, the results reveal a significant Berry curvature concentrated near the band-edge Kramers-Weyl point in $P2_1$-$BiAs_2Cl$, with an inhomogeneous distribution in momentum space. This inhomogeneity suggests the potential for spontaneous Berry curvature polarization. We computed the corresponding BCD tensor $D_a(\mathbf{K})$ and present its non-zero components in Figs 5c. As evident in the figure, when the Fermi level is tuned into the conduction band (as depicted in the upper panel of Figs S3.c), the $D_{zx}$ component, representing the gradient of $\Omega_x$ along $\mathbf{K}_z$, becomes prominent.[34] This suggests the potential for observing an intrinsic BCD-induced nonlinear Hall current in $P2_1$-$BiAs_2Cl$.

BCD induced nonlinear Hall effect is the current response to a static electric field. For finite frequency, there is another type of second order current response which results from the quantum metric (QM) contribution.[40-42] From Eq. 1 in Methods we can see that although both BCD and QM term originate from the Fermi surface, only the QM term contributes to the finite frequency response as it has pole structure $\partial_{k_\gamma} f_{mnk}/(\omega - \omega_{nmk})$. We further calculated the QM contribution to the nonlinear Hall current in electron-doped $P2_1$-$BiAs_2Cl$, as shown in Figs 5d. (see Methods for computational details). The computational results demonstrate that the QM contribution of Fermi surface at the band-edge Kramers-Weyl point constitutes a significant source of the nonlinear Hall current in electron-doped $P2_1$-$BiAs_2Cl$ at finite frequencies. It originates from the interband transition between the two SOC-split bands near the Kramers-Weyl point under the condition that one of the bands is cut by Fermi surface (see Fig. S3.c in Supplementary Materials). This contrasts with conventional transition metal dichalcogenide (TMD) systems, where the BCD typically dominates.[37-38] As shown in Fig. S7, the nonlinear Hall current from the BCD contribution at the same Fermi level (Fig. S3.c) is smaller than that from the QM contribution. This comparison clearly demonstrates the dominance of the QM mechanism in $P2_1$-$BiAs_2Cl$. This distinct mechanism further emphasizes $P2_1$-$BiAs_2Cl$ as a prime chiral material candidate for exploring intrinsic nonlinear Hall responses. Meanwhile, the strong nonlinear response below 0.2 eV indicates its application potential for infrared to THz range photon detections.

To identify additional candidate materials, we extended our study to other non-magnetic chiral semiconductors with high thermodynamic stability ($E_{hull} \leq 50$ meV/atom) and computed their band structures including SOC, as shown in Figs. S8-S11. Notably, although most of these materials exhibit "dirty" band structures and show minimal band splitting near the Weyl points compared to $BiAs_2Cl$, we still identified several structures featuring Weyl points near the band edges (Fig. S8). This finding suggests that these systems are also promising candidates for exploring the BCD and QM responses.

## B. *P*2$_1$3-$Pd_3SbB$: A Topological Chiral Material Featuring Six-fold Degenerate Band Crossing and Large Magnetoresistance

The structure configuration and Brillouin zone of $P2_13$-$Pd_3SbB$ was displayed in

Figs 2e. Compared to common Dirac or Weyl semimetals, the multiple degeneracy points in higher-fold degenerate semimetals (HFSMs), which can host distinctive massless fermions with higher degeneracy, afford electrons in their vicinity enhanced degrees of freedom for excitation. This characteristic provides a promising platform for exploring quantum phenomena and developing next-generation electronic devices. Chiral non-magnetic crystals constitute an excellent system for investigating higher-fold degeneracy. For instance, the CoSi-family represents a prominent class of chiral non-magnetic crystals.[7-8] Owing to constraints imposed by crystal symmetry, they exhibit an enforced six-fold degenerate node at the R point, rendering them an exceptional platform for exploring mixed-parity pairing and topological superconductivity.[43] Based on rigorous screening of the $E_{hull}$ and restrictions imposed by space group, we identified four thermodynamically favorable candidate compounds predicted for $M_xM'_yB_z$ compounds that potentially host higher-fold degeneracy: $Pt_3BiB$, $Pd_3SbB$, and $Pt_3GeB$ in space group $P2_13$, and $Pt_3Cd_2B$ in space group $P4_332$. We calculated their electronic band structures including SOC, with the results presented in Figs 6a and Figs S5.e-f. Analysis of the calculated band structures reveals that $P2_13$-$Pd_3SbB$ possesses a six-fold degenerate point at the R point very close to the Fermi level, while the degenerate points in the other systems lie significantly farther from the Fermi level. This indicates that $P2_13$-$Pd_3SbB$ is the most promising candidate material among these for probing six-fold degenerate fermions. Furthermore, these band structure calculations further confirm that chiral topological semimetals are excellent hosts for Kramers-Weyl fermions.

Using maximally localized Wannier functions (fitted results shown in Figs S3.e), we calculated the (001)-projected surface states of $P2_13$-$Pd_3SbB$ (Figs 6b) and the surface spectral function fixed at -0.1 eV below the Fermi level (Figs 6c). $P2_13$-$Pd_3SbB$ exhibits long Fermi arcs, similar to those observed in other chiral topological semimetals such as RhGe.[43] Furthermore, the calculated surface states show a clear gap from the bulk states, indicating that these Fermi arcs are promising candidates for experimental observation via angle-resolved photoemission spectroscopy (ARPES).[44] Additionally, given the presence of multiple topological band crossings near the Fermi surface of $P2_13$-$Pd_3SbB$, we investigated the influence of magnetic field on its resistivity within the framework of Boltzmann transport theory under the relaxation-time approximation.[45] First, we calculated the Hall resistivity ($\rho_{yx}$) as a function of magnetic field at different temperatures, as shown in Figs 6d. The results reveal a negative linear dependence of $\rho_{yx}$ on the magnetic field, analogous to that reported for the chiral topological semimetal CoSi family.[46] Subsequently, we calculated the longitudinal resistivity ($\rho_{xx}$) of $P2_13$-$Pd_3SbB$ along the $x$-direction (Figs 6e). It can be seen that $\rho_{xx}$ increases with temperature both with and without an applied magnetic field, mirroring the trend observed in CoSi and BeAu.[47-48] We further quantified the transverse magnetoresistance (MR) as a function of magnetic field and temperature. The MR was calculated using the formula $MR(\%) = [\rho_{xx}(B) - \rho_{xx}(0)] / \rho_{xx}(0) \times 100$. The results demonstrate that the MR of $P2_13$-$Pd_3SbB$ increases with increasing magnetic field strength. At 4 T, the MR reaches 150%. By comparison, CoSi exhibits an MR of 115% at 2 K.[46] Therefore, the MR in this compound is significantly higher than that in CoSi. Figs 6e also reveals that above 30 K, the MR of $P2_13$-$Pd_3SbB$ exhibits weak temperature dependence. Within the temperature range of 90–300 K, the MR value under a 10 T magnetic field remains approximately 225%.

Moreover, as mentioned previously, $P2_13$-$Pd_3SbB$ also exhibits superconductivity. Although its superconducting transition temperature is relatively low, its unconventional electronic structure, coupled with robust long Fermi-arc features, establishes this material as a candidate system for exploring hybrid pairing mechanisms and topological superconducting states.

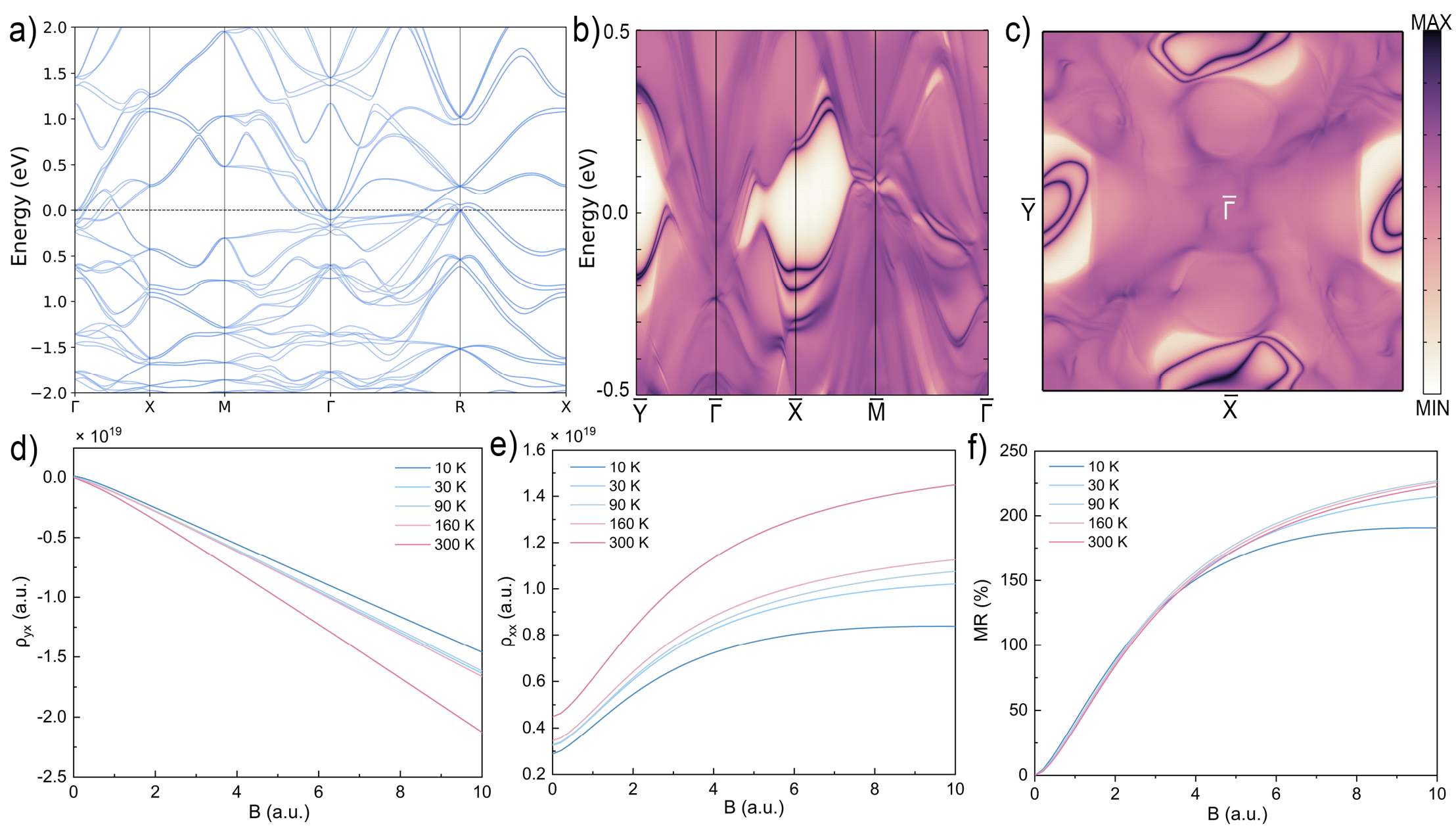

**Fig. 6 Band structures and magnetoresistance of *P*2₁3-Pd₃SbB.** a) Band structure calculated with SOC, b) The surface band spectrum projected on (001) surface, c) the surface band spectra at -0.1 eV below the Fermi energy, d) Hall resistivity ($\rho_{yx}$), e) transverse resistivity ($\rho_{xx}$ ), and (f) transverse magnetoresistance as functions of the magnetic field.

## Discussion

Driven by advances in computational materials science, the exploration of the vast unknown materials space for crystals with desirable properties has become a core research goal.[49-51] However, traditional DFT-based crystal structure prediction (CSP) methods face significant computational bottlenecks when applied to systems for large-scale, multi-elemental systems due to their immense computational cost. Encouragingly, the development of UMLPs, with their continuously improving accuracy and efficiency, is effectively bridging this gap.[15, 52-53] These meticulously trained foundation models reduce the computational cost of structure optimization and energy evaluation within CSP by several orders of magnitude while approaching DFT-level accuracy, thereby providing crucial support for high-throughput calculation workflows.

Unlike optimization algorithms that rely on specific initial configurations or preset symmetries, or generative models requiring specific training data, the Random Structure Search (RSS) method, a famous package is AIRSS,[54] necessitates no prior assumptions or constraints regarding any features of the target structure. By performing large-scale random sampling of atomic positions, elemental types, and stoichiometries within chemical space, it provides a powerful tool for exploring unknown, even "counterintuitive" structural regions. This approach holds promise for revealing metastable phases that conventional methods might overlook. Admittedly, for specific systems with relatively fixed components and chemical environments, physics-inspired optimization algorithms, such as particle swarm optimization in CALYPSO,[55] evolutionary algorithms in USPEX[56] and XtalOpt,[57] or data-driven generative models might offer advantages in search efficiency and precision.[58] However, when prediction tasks involve broad combinations of elements spanning the periodic table and highly variable compositional ratios, leading to a dramatic increase in the system's degrees of freedom, RSS often becomes a more indispensable choice due to its inherent universality and unbiased nature. The integrated prediction workflow developed in this study combine large-scale random structure generation and optimization using uMLIPs, and precise DFT validation is specifically aimed at addressing the challenge of exploring such multi-elemental materials.

When assessing the intrinsic stability of predicted crystal structures without external pressure, convex hull analysis and phonon calculations constitute the two fundamental pillars for determining thermodynamic and dynamical stability. Currently, most high-throughput computational workflows tend to predict $E_{hull}$ against known phases in large databases like the MP database. This approach identifies structures that are energetically stable relative to known phases, serving as a core screening step for candidate crystals.[49] However, a significant limitation arises from the frequent selective omission of computationally expensive phonon dispersion calculations within the workflow, or their restriction to the final validation of only a few select candidates. This compromise risks misclassifying a substantial number of dynamically unstable structures, i.e., those exhibiting imaginary frequencies in their phonon spectra, as stable, solely because they lie on or near the energy convex hull.

Encouragingly, a series of recent benchmark studies demonstrated that uMLIPs have achieved accuracy highly comparable to reference DFT calculations in identifying the presence or absence of imaginary frequencies to evaluate dynamical stability.[22] More significantly, the computing cost of phonon spectra with uMLIPs is a staggering three to six orders of magnitude lower than with DFT.[53] This dramatic reduction provides a realistic possibility for the large-scale, high-throughput integration of phonon stability screening into structure prediction workflows. It is important to note, however, that current uMLIPs generally tend to underestimate the lattice vibrational frequencies in phonon spectra. This systematic bias may impact the precise calculation of thermodynamic properties dependent on full spectral details, such as lattice thermal conductivity. Our findings (as shown in Figs. S12 a, b) clearly corroborate this consensus, for two representative materials discussed above, the phonon spectra calculated using uMLIPs agree with DFT results in determining the dynamical stability, but exhibit an observable systematic underestimation in the absolute frequency values especially for optical phonon branches.[59] Furthermore, as can be observed from Figs S12 c, d, for structures exhibiting phonon spectrum instability, although the discrepancy with DFT calculations is considerable in terms of imaginary frequency spectra for some configurations, the qualitative trends remain consistent.

Recognizing the inherent risk of relying solely on $E_{hull}$ and leveraging the demonstrated reliability of uMLIPs for qualitative assessment of phonon stability, this study implements high-throughput phonon dispersion calculations within its prediction workflow. Capitalizing on the speed advantage of uMLIPs, we systematically computed phonon spectra for the large pool of candidate structures that had passed the initial convex hull screening and structural optimization. This step is crucial, as it effectively identifies and filters out pseudo-stable crystals, i.e., energetically near the convex hull but exhibit imaginary frequencies in their phonon spectra. The phonon calculation results for some predicted compounds in this work are presented in Figs. S12. This practice of integrating high-throughput phonon validation significantly enhances the dynamical reliability of the final predictions, preventing potential false discoveries that could arise from omitting phonon calculations. It is also essential to note that within the current stage of the prediction workflow, the final verification and refinement of the uMLIPs preliminary results using DFT remains an indispensable and critical step. However, we are confident that with ongoing advancements in uMLIP architectures, training strategies, and data coverage, their accuracy and universality will steadily advance. Concurrently, the potential of the RSS method for discovering materials with unconventional structures and exceptional physical properties will be significantly amplified.

## Conclusion

In summary, this study implemented a high-throughput prediction workflow that integrates universal machine learning interatomic potentials (uMLIPs) with the Random Structure Search (RSS) method. This approach allowed for the large-scale exploration of multi-elemental systems. Through rigorous screening based on phonon stability, we have identified over 260 chiral inorganic crystals with an $E_{hull} \leq 0.2$ eV/atom. Among these, more than 60 exhibit excellent thermodynamic stability with $E_{hull} \leq 0.05$ eV/atom, making them particularly promising candidates for experimental systhesis. Particularly noteworthy are $P2_1$-$BiAs_2Cl$ and $P2_13$-$Pd_3SbB$. $P2_1$-$BiAs_2Cl$ exhibits significant potential for exhibiting the intrinsic nonlinear Hall effect driven by contributions from both the Berry curvature dipole (BCD) and quantum metric (QM).

$P2_13$-$Pd_3SbB$ exhibits a symmetry-enforced six-fold degenerate point near the Fermi surface. Its surface states display long Fermi arcs traversing the Brillouin zone, clearly separated from the bulk states and exhibits a large magnetoresistance. This work expands the pool of candidate chiral topological inorganic materials and demonstrates a symmetry-constrained framework for large-scale, low-cost crystal structure prediction in such systems.

## Methods

***Crystal Structure Prediction***. The high-throughput symmetry constrained crystal structure generation capability, described in this paper, is an extension based on the basic functionality of PyXtal package.[19] We selected mattersim-v1.0.0-1M as the foundation model for structure optimization and mattersim-v1.0.0-5M for phonon calculations.[20] Structural optimization, crystal structure processing, and symmetry determination were performed using atomic simulation environment (ASE),[60] pymatgen,[61] and spglib,[62] with a symmetry tolerance of 0.05 Å for space group identification.

***DFT calculation***. $E_{hull}$ related calculations were performed with the Vienna ab initio Simulation Package (VASP),[63] employing the PBE exchange-correlation functional within the generalized gradient approximation (GGA).[64] A plane-wave energy cutoff of 520 eV was used, and the force convergence criterion was set to 0.02 eV/Å. Brillouin-zone integration was performed using Γ-centered k-point meshes with a spacing of $2\pi\times0.03$ Å$^{-1}$.

Property calculations via DFT were primarily conducted using Quantum Espresso,[65-66] employing the ONCV pseudopotential.[67] Fitting of maximally localized Wannier functions was achieved using Quantum Espresso and WANNIER90 package.[68] For the $P2_1$-$BiAs_2Cl$ system, the k-mesh of Brillouin zone sampling was set to 8×8×6, the Bi–p, As-p and Cl-p orbitals were utilized for the initial projection. For the $P2_13$-$Pd_3SbB$ system, the k-mesh was set to 7×7×7, the s, d orbitals of Pd were utilized for the initial projection.

***Nonlinear Optical Property.*** The second harmonic generation (SHG) coefficients for insulators were calculated using the expression[69-71]

$$\chi^{(2)}_{\alpha\beta\gamma}(2\omega;\omega,\omega) = \frac{-ie^3}{2m^3\tilde{\omega}^3 V}\sum\nolimits_{nml\mathbf{k}} \left[ \begin{array}{c} \frac{p^{\alpha}_{nmk}p^{\beta}_{mlk}p^{\gamma}_{lnk}f_{nlk}}{(2\tilde{\omega}-\omega_{mnk})(\tilde{\omega}-\omega_{lnk})} + \frac{p^{\alpha}_{nmk}p^{\gamma}_{mlk}p^{\beta}_{lnk}f_{mlk}}{(2\tilde{\omega}-\omega_{mnk})(\tilde{\omega}-\omega_{mlk})} \\ + \frac{p^{\alpha}_{nmk}p^{\gamma}_{mlk}p^{\beta}_{lnk}f_{nlk}}{(2\tilde{\omega}-\omega_{mnk})(\tilde{\omega}-\omega_{lnk})} + \frac{p^{\alpha}_{nmk}p^{\beta}_{mlk}p^{\gamma}_{lnk}f_{mlk}}{(2\tilde{\omega}-\omega_{mnk})(\tilde{\omega}-\omega_{mlk})} \end{array} \right] \quad (1)$$

where $p^{\alpha}_{nmk} = \langle nk|p^{\alpha}|mk\rangle$ denotes the matrix element of momentum operator, $\tilde{\omega} = \omega + i\eta$, $\eta$ arises from the adiabatic switching-on process which is fixed to 0.01 eV to represent the finite lifetime of electron-hole pairs, $V$ is the crystal volume, $f_{mnk} = f_{mk} - f_{nk}$ and $\omega_{mnk} = E_{mk} - E_{nk}$ stand for the occupation and energy difference between band n and m. Therefore the SHG coefficients can be directly calculated from the momentum matrix element and Kohn-Sham single particle energies from DFT calculation. In Table I, the Γ-centered k-mesh is spaced at 2π × 0.025 $Å^{-1}$. The total second order conductivity for metallic system or semiconductor with electron or hole doping were calculated using the expression[41-42, 72-75]

$$\sigma^{(2)}_{\alpha\beta\gamma}(0;-\omega,\omega) = \sigma^{(2),SC}_{\alpha\beta\gamma} + \sigma^{(2),BCD}_{\alpha\beta\gamma} + \sigma^{(2),QM}_{\alpha\beta\gamma} =$$

$$-\frac{i\pi e^3}{2\hbar^2 V}\sum\nolimits_{nmk} f_{nmk}(r^{\beta}_{mnk}r^{\gamma}_{nmk;\alpha} + r^{\gamma}_{mnk}r^{\beta}_{nmk;\alpha})\delta(\omega - \omega_{nmk}) +$$

$$\frac{e^3}{2\hbar^2 V}\sum_{nmk}\left[\frac{A^{\alpha}_{mnk}A^{\gamma}_{nmk}\,\partial_{k_\beta} f_{mnk}}{\widetilde{\omega}}+\frac{A^{\alpha}_{mnk}A^{\beta}_{nmk}\,\partial_{k_\gamma} f_{mnk}}{(\widetilde{\omega}-\omega_{nmk})}\right] \quad (2)$$

where $\sigma^{(2),SC}_{\alpha\beta\gamma}$ is the shift current which contributes to the bulk photovoltaic effect,[76-82] $\sigma^{(2),BCD}_{\alpha\beta\gamma}$ is the Berry curvature dipole (BCD) and quantum metric (QM) contribution originating from the intraband transition on the Fermi surface, which results into the nonlinear Hall effect.[34, 37-38, 42] In the above expression, the Berry connection $A^{\alpha}_{nmk}$ is related to the r-matrix,

$$A^{\alpha}_{nmk} = i\left\langle nk\middle|\frac{\partial}{\partial k^{\alpha}}\middle|mk\right\rangle = r^{\alpha}_{nmk} \quad (3)$$

and the generalized derivative reads,[68]

$$r^{\beta}_{nmk;\alpha} = \frac{\partial r^{\beta}_{nmk}}{\partial k^{\alpha}} - i r^{\beta}_{nmk}(A^{\beta}_{nnk} - A^{\beta}_{mmk}) \quad (4)$$

The k-mesh for $\sigma^{(2),SC}_{\alpha\beta\gamma}$, BCD and $\sigma^{(2),QM}_{\alpha\beta\gamma}$ results in Fig. 5 are 100×100×100 (including 48 bands, see Figs S3. d), 72×72×48, 36×36×24, respectively. Different from the calculation of $\chi^{(2)}_{\alpha\beta\gamma}(2\omega;\omega,\omega)$ the expression of $\sigma^{(2)}_{\alpha\beta\gamma}(0;-\omega,\omega)$ involves k-space derivative of occupation function which represents the Fermi surface contribution. We use the Lorentzian approximation of Dirac-$\delta$ function where we can see that $\partial_{k_\alpha} f_{nk}$ has pole on the Fermi surface $E_F$,

$$\partial_{k_\alpha} f_{nk} = -\frac{\hbar}{m} p^{\alpha}_{nnk}\,\delta(E_{nk}-E_F) \cong -\frac{\hbar}{m} p^{\alpha}_{nnk}\frac{1}{\pi}\frac{\tau}{(E_{nk}-E_F)^2+\tau^2} \quad (5)$$

where the broadening parameter $\tau$ is chosen to be 0.08 eV, which is able to reproduce the BCD on a much finer k-mesh of 72×72×48.

***Superconductivity.*** In electron-phonon coupling (EPC) strength calculations. The plane wave basis energy cutoff was set to 50-70 Ry, and the Brillouin zone was sampled using Γ-centered k-point meshes with 12×12×12 or 16×16×16, the q-point mesh was set to 2×2×2 for rough screening and 4×4×4 for accurate evaluation. The superconducting transition temperatures ($T$c) were calculated from Allen−Dynes equation.[83-84]

$$T_{\mathrm{c}} = \omega_{log}\frac{f_1 f_2}{1.2} exp\left[\frac{-1.04\cdot(1+\lambda)}{\lambda-\mu^*(1+0.62\lambda)}\right] \quad (6)$$

$$f_1 f_2 = \left[1+\left(\frac{\lambda}{2.46\cdot(1+3.8\mu^*)}\right)^{\frac{3}{2}}\right]^{\frac{1}{3}} \cdot \left[1-\frac{\left(1-\frac{\omega_2}{\omega_{log}}\right)\lambda^2}{3.312\cdot(1+6.3\mu^*)^2+\lambda^2}\right] \quad (7)$$

where $\omega_{\log}$ stands for the logarithmic average frequency and $\omega_2$ is the mean-square frequency. The μ* was set to 0.1 and the EPC constant (λ) can be defined as

$$\lambda = \sum_{qv} \lambda_{qv} = \int \frac{2\alpha^2 F(\omega)}{\omega} d\omega \quad (8)$$

Phonon calculations were carried out with the Phonopy package.[85] For uMLIP-based and DFT-based phonon calculations, 2×2×2 supercells and an atomic displacement of 0.02 Å were used. The topological characteristics were calculated with WannierTools and Wannierberri package.[86-87] In addition, PyProcar[88] and VESTA[89] code was also performed in visualization and figure plotting.

## Data availability

The crystal structure files and phonon dispersion results supporting the findings of this study are available in the Zenodo repository with the DOI: 10.5281/zenodo.17374067.

## Author’s contributions

**Jiexi Song**: calculation, data analysis, preparation of paper. **Diwei Shi**: discussion, data analysis. **Fengyuan Xuan**: discussion, paper editing and review, funding acquisition. **Chongde Cao**: discussion, paper editing and review, supervision, funding acquisition.

## Conflicts of interest

There are no conflicts of interest to declare.

## Acknowledgements

This work was supported in part by the National Science and Technology Major Project (Grants No. 2023ZD0120702), the Space Application System China Manned Space Program (ZDBS-ZRKJZ-TLC021), the National Natural Science Foundation of China (52271037), the Shaanxi Provincial Natural Science Fundamental Research Program, China (2023-JC-ZD-23), the Natural Science Foundation of Jiangsu Province (Grant No. BK20240395), the Jiangsu Funding Program for Excellent Postdoctoral Talent (Grant No. 2025ZB701) and Opening Grant of Zhejiang Key Laboratory of Data-Driven High-Safety Energy Materials and Applications (OG2024008). Calculations were performed on Sugon HPC clusters equipped with HYGON X86 32-core processors (2.5 GHz) and at the Beijing Super Cloud Computing Center.

## Safety Statement

This work involves only theoretical and computational studies. No experimental procedures, hazardous materials, or safety risks are associated with the reported research.

# Supplementary Materials

## Accelerating Discovery of Ternary Chiral Materials via Large-Scale Random Crystal Structure Prediction

Jiexi Song [a, #], Diwei Shi [b, #], Fengyuan Xuan [a,*] and Chongde Cao [c,**]

a Suzhou Laboratory, Suzhou 215123, China
b School of Naval Architecture and Maritime, Zhejiang Ocean University, Zhoushan 316022, China
c School of Physical Science and Technology, Northwestern Polytechnical University, Xian 710072, China

#: Jiexi Song and Diwei Shi contributed equally

Author to whom correspondence should be addressed:
xuanfy@szlab.ac.cn; caocd@nwpu.edu.cn

This Supplementary Materials contains screening results on $TM_xM_yCh_z$ systems, the predicted compounds list in $M_xX_yHal_z$ and $M_xM'_yB_z$ systems, and the electronic properties, SHG coefficients, band fitting results of some representative materials. .

### Details of $E_{hull}$ calculation

The calculation of formation energies and energies above hull proceeded as follows, using the Materials Project (MP) database as the reference for all known competing phases:

1. Reference and Correction Scheme: The known elements and compounds in MP served as our reference set. We adopted the MP gas-phase energy correction (Ref.[1]) to address significant DFT errors for systems containing the following elements: S, F, Cl, Br, I, N, H, Se, Si, Sb, and Te.

2. Implementation for DFT: For density functional theory (DFT) calculations, this correction was applied by directly using the MaterialsProject2020Compatibility() scheme.

3. Implementation for uMLIP: For energies obtained from the uMLIP potential, we developed a class that emulates the specific pseudopotentials used in MP's DFT calculations. This allowed us to apply an equivalent gas-phase correction, yielding consistent and corrected energies above hull.

**Comparison to existing databases**

Under the specified constraints for elemental combinations and the number of atoms per unit cell ($\leq$ 24 atoms), we identified a structural type consistent with experimentally synthesized compounds, specifically, the $AHalO_3$ family. As illustrated in Figure S1.c, the predicted "parent" structure $NaBrO_3$ was compared with its experimentally reported counterpart in the Materials Project (MP) database. Following this identification, through systematic substitution and screening of elements within the same group, additional compounds such as $NaClO_3$ and $NaIO_3$, which adopt the $P2_13$ space group, were successfully selected. This outcome validates the capability of our approach to recover some known crystal structures.

Furthermore, we encountered another notable example: $NbI_2O$, a composition already reported in the MP database as an experimentally synthesized quasi-two-dimensional chiral crystal with space group $C2$. Upon structural relaxation, our prediction for $NbI_2O$ yielded a structure with the $C2/m$ space group. The discrepancy between our predicted structure and the experimental one arises from differences in the relative stacking positions of the two-dimensional layers, which result in the loss of chirality, leading to its exclusion during HT-screening. Analysis of the DFT-calculated total energies reveals that the *C*2/*m* and *C*2 structures exhibit nearly identical energies, whereas other predicted $NbI_2O$ polymorphs are significantly higher in energy. However, due to symmetry-based filtering criteria applied during the selection process, the non-chiral *C*2/*m* structure was excluded. The absence of a subsequent elemental substitution step for this system also prevented the identification of another experimentally known compound, *C*2-$NbCl_2O$.

These observations suggest that relaxing symmetry constraints, for instance, by expanding from 65 Sohncke space groups to all 230 space groups could substantially increase the number of predicted structures near the ground-state (i.e., with $E_{hull} = 0$). For instance, we present several non-chiral structures predicted by our method that exhibit an $E_{hull} = 0$ in Fig.S1.d.

It is worth noting that, within the current MP database, for the compositional space $M_xX_yHal_z$ (with x, y, z = 1–3) and under the constraints of non-magnetic character, $E_{hull} = 0$, and chiral space group, only nine structures satisfy all criteria: $CsSeBr_3$ (mp-2014596), $NbI_2O$ (mp-549720), $NbCl_2O$ (mp-1025567), $NbI_3O$ (mp-546285), $NaBrO_3$ (mp-23339), ReSeCl (mp-676712), $NbBr_2O$ (mp-550070), $Ca_3PI_3$ (mp-27272), and $NbOF_3$ (mp-760762). For the $M_xM'_yB_z$ system, no structures simultaneously satisfying these conditions are present in the MP database.

Nevertheless, the effectiveness of our method in recovering “known structures” is directly influenced by the diversity and quantity of initially generated random structures.

A negative example from this work is the $TM_xM_yCh_z$ system, which contains several non-magnetic, chiral structures with $E_{hull}$ = 0, such as SbIrS (mp-8630), SnTePt (mp-1218898), and TeAsIr (mp-1217272). However, in our structural predictions for this system, no configurations with $E_{hull}$ = 0 were identified. A recent benchmark study (arXiv:2407.00733) indicated that the highest success rate among existing crystal structure prediction algorithms for reproducing known experimental structures is only 46.111%, underscoring the considerable challenges that remain in this field. For our work, the present results indicate that future improvements should focus on enhancing the randomness and diversity of initial structural sampling and on reducing the computational cost associated with large-scale structural optimization.

Furthermore, since the $E_{hull}$ in our calculations were already derived using reference phases from the Materials Project database, we primarily focus in Tables S1 and S3 on comparing the energy differences with those crystals extracted from the OQMD database.

**Benchmark**

(1) Benchmarking SHG coefficients for a known NLO compound identified in our screening

Among the noncentrosymmetric candidates identified by our high-throughput search, we examined the second-order nonlinear optical (NLO) response, the $\chi^{(2)}_{123}$ component (equivalent to $2d_{14}$), for compounds with existing literature data. As listed in Table S2, the most representative example is $P2_13$-$NaBrO_3$. Our computed $\chi^{(2)}_{123}$ value 3.9 pm/V shows good agreement with the previously reported theoretical result 3.0024 pm/V from the literature (*Chemical Physics Letters* 287 1998 503–508).

(2) Extended workflow validation on $LiGaS_2$

Noting that our original post-processing step retained only the lowest-energy structure per composition, we further tested our workflow on $LiGaS_2$, a NLO material with polymorphs. In this extended test, we deliberately increased the number of candidate structures extracted after the uMLIP-based relaxation, from just one to dozens to explore metastable phases more thoroughly.

This enriched sampling successfully recovered both reported noncentrosymmetric phases: $Pna2_1$ and $I\text{-}42d$. Moreover, we identified several additional low-energy polymorphs that are dynamically stable at the uMLIP level (Figure S13). Crucially, we computed the SHG coefficients for the $Pna2_1$ and $I\text{-}42d$ phases using DFT protocol and obtained values of $d_{15}$ = -5.12, $d_{24}$ =-4.68, $d_{33}$=13.04 pm/V for $Pna2_1$ phase and $d_{36}$=-9.75 pm/V for $I\text{-}42d$ phase. These results are also in close agreement with previously reported values $d_{15}$ = -5.33, $d_{24}$ =-4.93, $d_{33}$=12.28 for $Pna2_1$ phase and $d_{36}$=-10.28 pm/V for $I\text{-}42d$ phase (*ACS Appl. Mater. Interfaces* 2024, 16, 36658−36666).

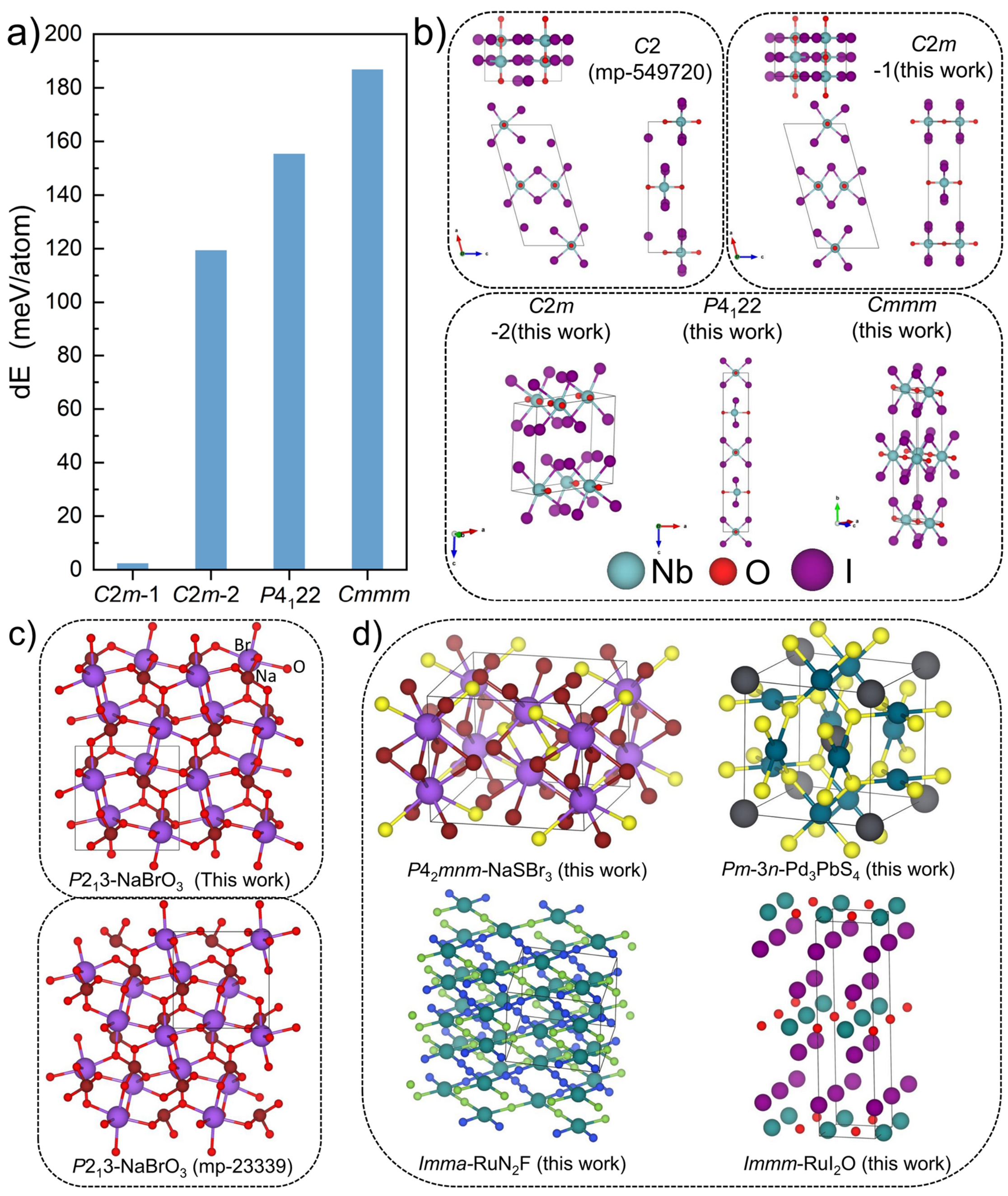

**Fig. S1** a) DFT energy differences of $NbI_2O$ predicted in this work with that in MP database, b) Crystal structures of $NbI_2O$, c) Crystal structures of $NaBrO_3$ in this work and MP database, d) Some predicted non-chiral crystal structures in this work.

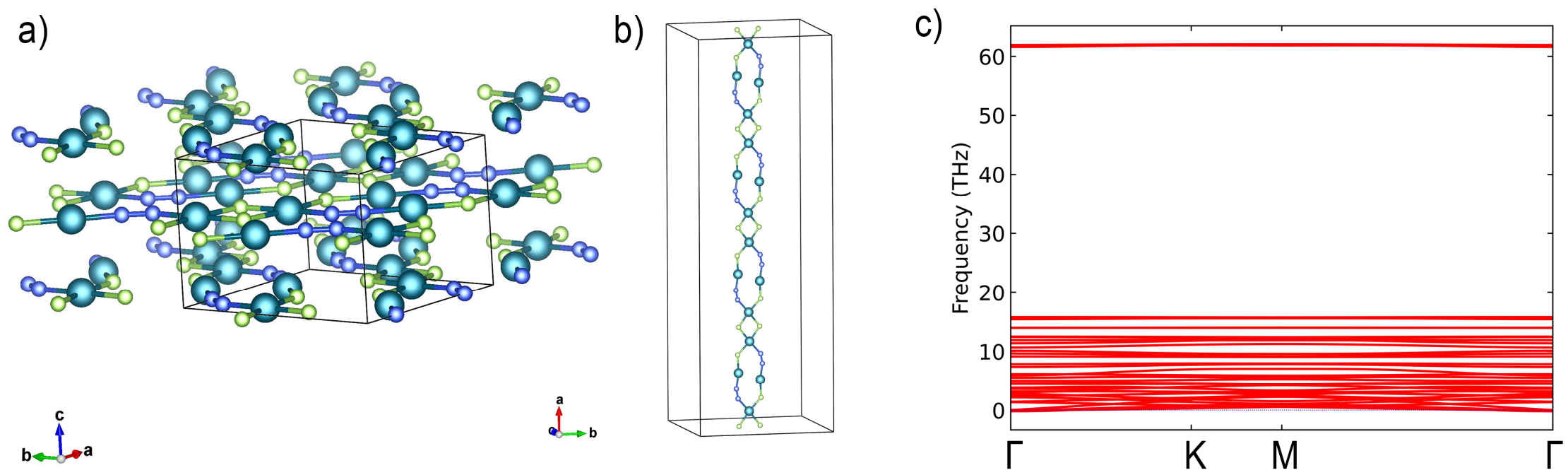

**Fig. S2 Overview of predicted compounds *Pnnm*-PdNF.** a) Structure image of bulk *Pnnm*-PdNF, b) the 1D chain of the -Pd-N-F-, c) Phonon spectrum of 1D-PdNF.

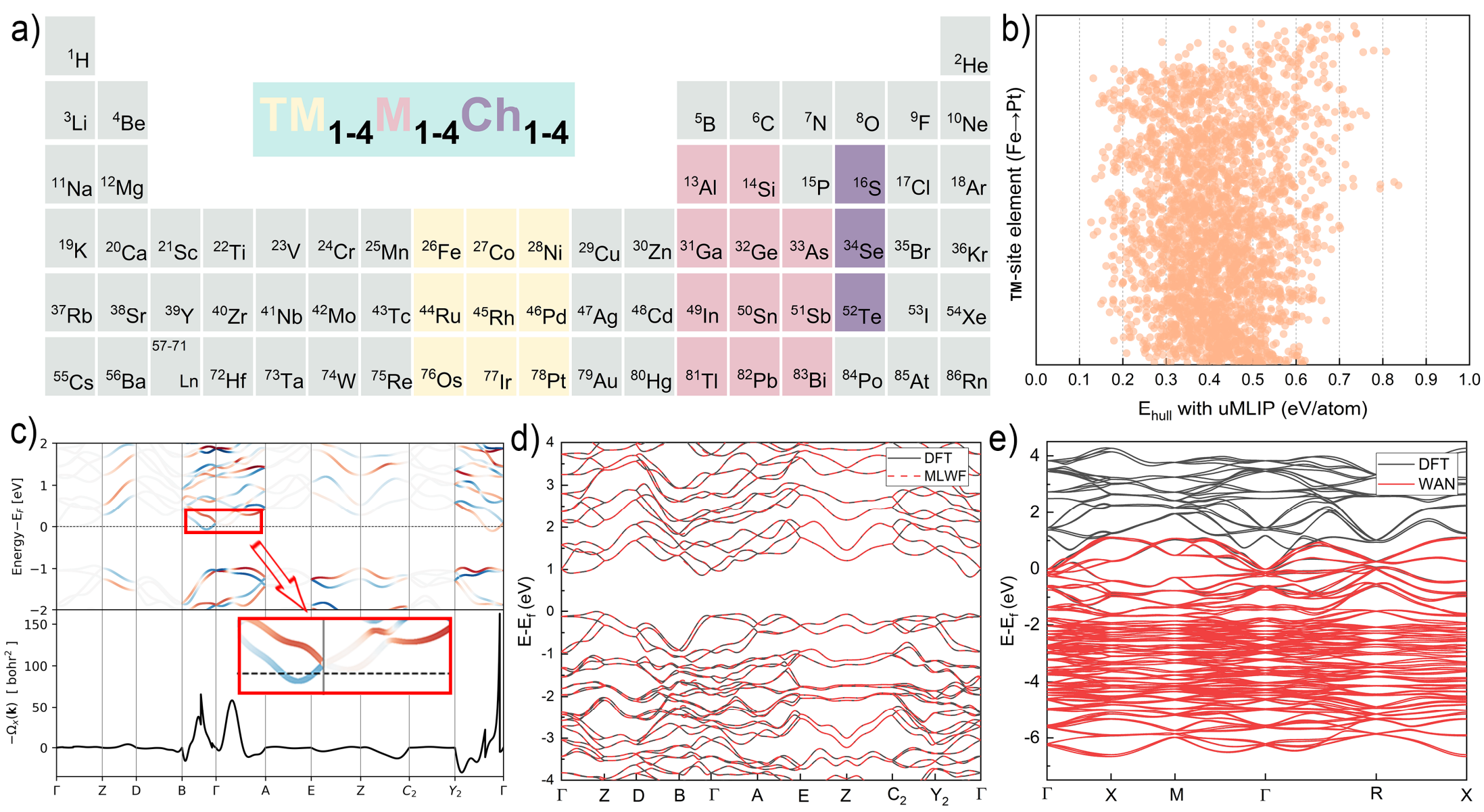

**Fig. S3 Screening results of $TM_xM_yCh_z$ system and fitting results for specific compounds.** a) Elemental selection for the $TM_xM_yCh_z$ system. b) Distribution of energies above convex hull ($E_{hull}$) for the lowest-energy structure extracted for each composition within the $TM_xM_yCh_z$ system. c) The spin-projected band structure of $P2_1$-$BiAs_2Cl$ with spin-orbit coupling (SOC) and the integrated berry curvature $\Omega_x$, the embedded illustration shows the position of the Fermi level. d) Calculated energy bands from GGA+PBE and MLWF of $P2_1$-$BiAs_2Cl$ with considering SOC. e) Calculated energy bands from GGA+PBE and MLWF of $P2_13$-$Pd_3SbB$ with considering SOC.

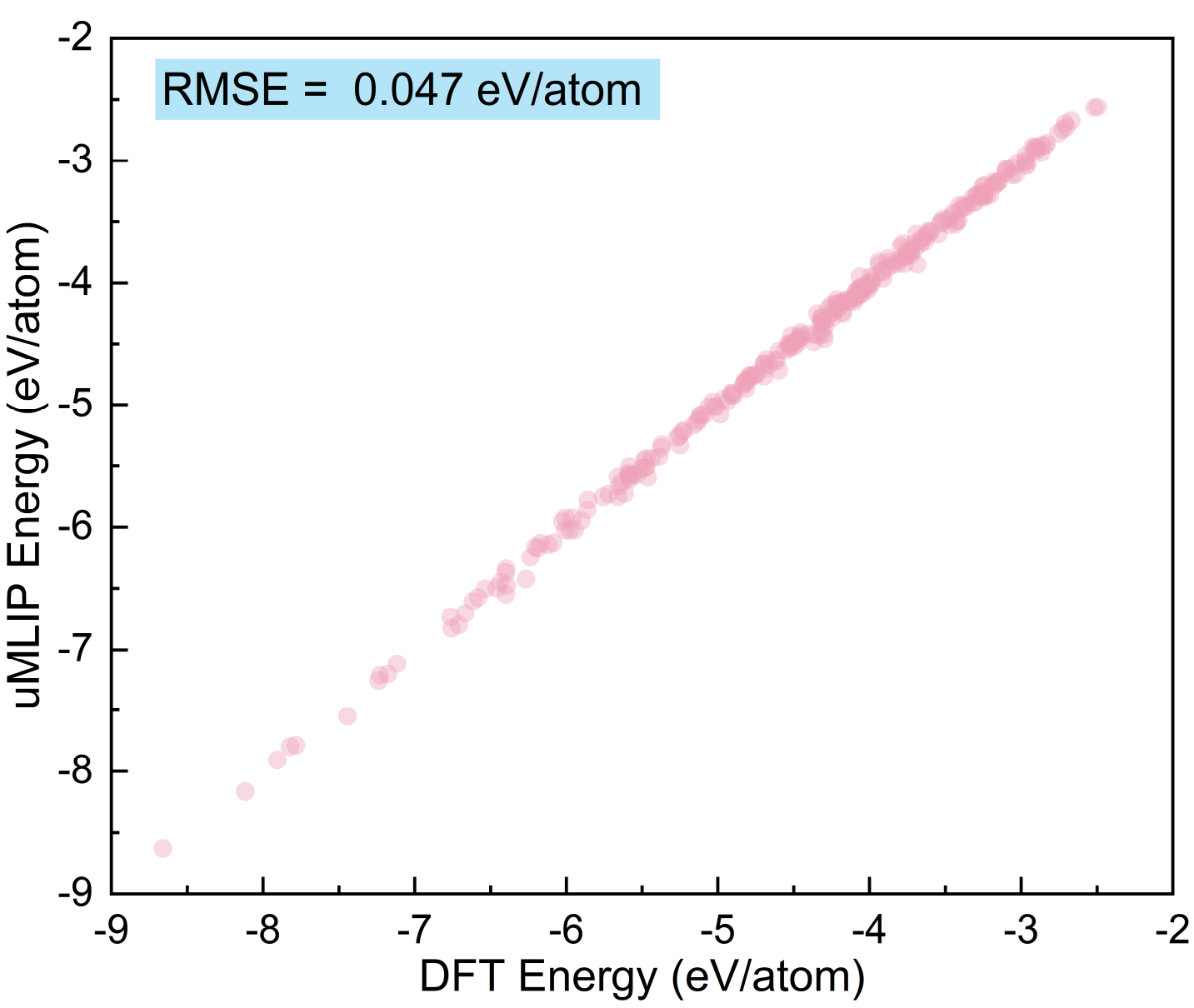

**Fig. S4 The energy comparisons between calculated DFT energy and uMLIP energy for predicted stable chiral structures in this work, the RMSE ($\sqrt{\frac{1}{N}\sum_{1}^{N}(E_{DFT}-E_{uMLIP})^2}$ is 0.047 eV/atom.**

**Table S1**. The $E_{hull}$ and PBE band gap of Predicted Compound in A=$M_xX_yHal_z$ system.

| | Metal (m) or band gap (eV) | $E_{hull_DFT,}$ ( eV) | Target compound exists in OQMD | Space Group In OQMD | δE, $E_{this_work}$-$E_{OQMD}$ (eV/atom) |
|---|---|---|---|---|---|
| $P2_12_12_1$-AlSe$_2$Cl | 1.43 | 0.00 | × | | |
| $P2_12_12_1$-BeS$_3$Cl$_2$ | 2.06 | 0.00 | × | | |
| $P2_12_12_1$-CaS$_3$Cl$_2$ | 1.78 | 0.00 | × | | |
| $P2_12_12_1$-PbS$_3$I | m | 0.00 | × | | |
| $P2_12_12_1$-TiSe$_2$I$_2$ | 0.57 | 0.00 | × | | |
| $P2_12_12_1$-ZrS$_2$Br$_2$ | 1.52 | 0.00 | √ | *P*-1 | 0.08 |
| $P2_12_12_1$-TlS$_3$Cl | 1.60 | 0.00 | × | | |
| $P2_13$-AgBrO$_3$ | 2.45 | 0.00 | √ | *R*3*c* | -0.03 |
| $P2_1$-TiSe$_3$Br$_2$ | 0.67 | 0.00 | × | | |
| $P2_1$-ZrSe$_3$Br$_2$ | 0.75 | 0.00 | × | | |
| $P3_112$-RuO$_2$F | m | 0.00 | √ | *Cc* | -0.01 |
| $P3_1$-CuN$_2$Br | 0.79 | 0.00 | × | | |
| $P3_1$-CuN$_2$Cl | 0.53 | 0.00 | × | | |
| $P3_1$-InS$_3$Br$_3$ | 1.72 | 0.00 | × | | |
| $P3_1$-InS$_3$I$_3$ | 1.26 | 0.00 | × | | |
| $P3_2$-ReN$_3$Br | m | 0.00 | × | | |
| $P3_2$-ReN$_3$Cl | m | 0.00 | × | | |
| $P4_1$-ReNF$_3$ | m | 0.00 | × | | |
| $P4_222$-$IrN_2F_2$ | m | 0.00 | × | | |
| $P6_3$-$CdSe_3Br_2$ | 0.71 | 0.00 | × | | |
| $P6_522$-$CoS_2Br$ | 0.41 | 0.00 | × | | |
| $P6_522$-$RhS_2Cl$ | 0.75 | 0.00 | × | | |
| $Pnnm$-PdNF | m | 0.00 | √ | $C2/m$ | -0.01 |
| $P2_12_12_1$-$AlSe_2Br$ | 1.59 | 0.01 | × | | |
| $P2_12_12_1$-$HfSe_2Br_2$ | 1.52 | 0.01 | √ | $P$-1 | 0.04 |
| $P2_12_12_1$-$PbSe_3I$ | m | 0.01 | × | | |
| $P2_13$-$KBrO_3$ | 4.66 | 0.01 | √ | $R3m$ | -0.02 |
| $P2_13$-$NaBrO_3$ | 4.45 | 0.01 | √ | $P2_13$ | -0.01 |
| $P2_13$-$NaIO_3$ | 4.07 | 0.01 | √ | $Pca2_1$ | -0.01 |
| $P2_1$-$BiAs_2Cl$ | 0.35 | 0.01 | × | | |
| $P3_121$-$AuSe_2Cl$ | m | 0.01 | √ | $P2_1/c$ | 0.04 |
| $P2_12_12_1$-$ZrSe_2I_2$ | 0.93 | 0.03 | √ | $P$-1 | 0.09 |
| $P2_13$-$CsBrO_3$ | 4.67 | 0.03 | √ | $R3m$ | 0.01 |
| $P2_13$-YSF | 3.89 | 0.03 | √ | $P4/nmm$ | 0.01 |
| $P2_1$-$BiAs_2Br$ | m | 0.03 | × | | |
| $P2_1$-$InSCl_2$ | 1.45 | 0.03 | × | | |
| $P2_1$-$InS_3F_3$ | 1.46 | 0.03 | × | | |
| $P3_221$-$InS_3I$ | m | 0.03 | × | | |
| $P4_122$-$BiS_2F_3$ | 2.07 | 0.03 | × | | |
| $P4_1$-$AlS_3F$ | 2.40 | 0.03 | × | | |
| $P4_3$-$GeS_3Br_2$ | 1.33 | 0.03 | × | | |
| $P6_522$-$RhS_2Br$ | 0.83 | 0.03 | × | | |
| $P2_12_12_1$-$HfS_2Br_2$ | 1.74 | 0.04 | √ | $P$-1 | 0.08 |
| $P2_1$-$SbAs_2Br$ | m | 0.04 | × | | |
| $P4_32_12$-$TlS_2I$ | 1.43 | 0.04 | √ | $I4/mcm$ | -0.48 |
| $C2$-SrIN | 0.87 | 0.05 | × | | |
| $P2_12_12_1$-$CaS_3F_2$ | 2.25 | 0.05 | × | | |
| $P2_12_12_1$-$HfSe_2I_2$ | 1.13 | 0.05 | √ | $C2/m$ | 0.07 |
| $P2_13$-$AgIO_3$ | 2.54 | 0.05 | √ | $Pca2_1$ | 0.01 |
| $P2_13$-$NaClO_3$ | 5.58 | 0.05 | √ | $P2_13$ | 0.00 |
| $P2_1$-$TlS_3F_3$ | 0.50 | 0.05 | × | | |
| $P3_2$-$SnSe_3I$ | m | 0.05 | × | | |
| $P4_3$-AlSCl | 3.84 | 0.05 | √ | $P$-1 | 0.06 |
| $P6_3$-$ZnS_3Cl_2$ | 0.08 | 0.05 | × | | |
| $P2_12_12_1$-$PbS_2I_2$ | 1.08 | 0.06 | × | | |
| $P2_12_12_1$-SrSBr | 2.51 | 0.06 | × | | |
| $P2_12_12_1$-TlOF | 0.45 | 0.06 | √ | $Pbca$ | 0.02 |
| $P2_13$-YSeF | 3.42 | 0.06 | √ | $P2_1/m$ | 0.02 |
| $P2_1$-$CaS_3I_2$ | m | 0.06 | × | | |
| $P2_1$-$LaS_3F$ | 0.44 | 0.06 | × | | |
| $P2_1$-$PdSBr_3$ | 0.39 | 0.06 | × | | |
| $P3_121$-$RuI_3O$ | m | 0.06 | √ | $P1$ | 0.046 |
| $P3_1$-$MgN_3Cl$ | 3.75 | 0.06 | × | | |
| $P3_1$-$TlSeCl_2$ | 1.25 | 0.06 | × | | |
| $P3_2$-$CuS_3I$ | m | 0.06 | × | | |
| $P2_1$-$BiP_2Br$ | 0.79 | 0.07 | × | | |
| $P2_1$-$PdS_2Br_2$ | m | 0.07 | √ | $P$-1 | 0.20 |
| $P2_1$-$SbAs_2I$ | m | 0.07 | × | | |
| $P3_1$-$GaS_3I_3$ | 0.98 | 0.07 | × | | |
| $P4_12_12$-$NbIO_2$ | 0.33 | 0.07 | √ | $Pnma$ | 0.05 |
| $P4_1$-$CdS_2Br_2$ | 0.64 | 0.07 | × | | |
| $P4_1$-$ZrS_2Br_2$ | 1.72 | 0.07 | √ | $P$-1 | 0.16 |
| $P4_3$-$AlSeBr$ | 2.03 | 0.07 | √ | $C2/m$ | 0.12 |
| $P4_3$-$GeSe_2Cl_2$ | 0.92 | 0.07 | × | | |
| $P6_3$-$CdS_3Cl_2$ | m | 0.07 | × | | |
| $P2_12_12_1$-$GaSe_2Cl$ | 1.08 | 0.08 | × | | |
| $P3_1$-$ZrS_3I_2$ | 0.47 | 0.08 | × | | |
| $P3_221$-$ZrSe_2Cl_2$ | 1.57 | 0.08 | √ | $P$-1 | 0.13 |
| $P3_2$-$BaPI$ | 0.94 | 0.08 | × | | |
| $P3_2$-$SrPI$ | 1.02 | 0.08 | × | | |
| $P4_3$-$AlSeCl$ | 3.14 | 0.08 | √ | $C2/m$ | 0.11 |
| $P6_1$-$ReNF_2$ | 0.06 | 0.08 | × | | |
| $P6_522$-$CoS_2I$ | 0.46 | 0.08 | × | | |
| $R3$-$GeS_3Br_3$ | 0.01 | 0.08 | × | | |
| $P2_12_12_1$-$SrS_3Br_2$ | 1.19 | 0.09 | × | | |
| $P2_1$-$YSF_2$ | 2.50 | 0.09 | × | | |
| $P2_1$-$ZrS_3Br_2$ | 1.11 | 0.09 | × | | |
| $P2_1$-$ZrSF_3$ | 2.55 | 0.09 | × | | |
| $P3_121$-$OsI_3O$ | m | 0.09 | √ | $P4_2/mnm$ | 0.07 |
| $P3_1$-$AlSBr_2$ | 1.36 | 0.09 | × | | |
| $P3_1$-$NbS_3I_3$ | 0.86 | 0.09 | × | | |
| $P4_3$-$InS_2I$ | 0.89 | 0.09 | × | | |
| $P2_13$-$BaI_2O_3$ | 3.25 | 0.10 | × | | |
| $P2_13$-$LaSF$ | 3.46 | 0.10 | √ | $P4/nmm$ | 0.05 |
| $P2_1$-$PtN_3F$ | 0.09 | 0.10 | × | | |
| $P3_221$-$BaSeBr$ | 1.62 | 0.10 | × | | |
| $P3_2$-$CaSeI$ | 1.89 | 0.10 | × | | |
| $P4_322$-$InSCl$ | 1.87 | 0.10 | √ | $P2_1/c$ | 0.08 |
| $P4_3$-$BaSeI$ | 1.97 | 0.10 | × | | |
| $P2_12_12_1$-$GaS_2Cl$ | 0.75 | 0.11 | × | | |
| $P2_12_12_1$-$SbSeI$ | 0.62 | 0.11 | √ | $P2_1/c$ | 0.13 |
| $P2_12_12$-$InS_3Cl$ | 0.81 | 0.11 | × | | |
| $P2_12_12$-$OsSe_2Cl_2$ | m | 0.11 | × | | |
| $P3_2$-$GeSe_2Cl$ | 1.26 | 0.11 | × | | |
| $P2_12_12_1$-$AuSe_2Br$ | m | 0.12 | √ | $Cm$ | 0.20 |
| $P2_1$-$BiS_2I_2$ | 0.68 | 0.12 | × | | |
| $P3_112$-$CoSe_2I$ | m | 0.12 | × | | |
| $P3_121$-$GaS_2Br_2$ | m | 0.12 | × | | |
| $P3_1$-$TaS_2Cl_3$ | 1.29 | 0.12 | × | | |
| $P4_12_12$-$TaIO_2$ | 1.22 | 0.12 | √ | *Cmcm* | 0.12 |
| $P4_1$-BaSI | 2.19 | 0.12 | × | | |
| $P4_1$-$GaS_2I$ | 1.63 | 0.12 | × | | |
| $P4_1$-$NbS_3I_2$ | m | 0.12 | × | | |
| $P2_12_12_1$-$AuIO_3$ | 1.21 | 0.13 | √ | *R3c* | -0.13 |
| $P2_12_12_1$-$BiSe_2I_2$ | m | 0.13 | × | | |
| $P2_12_12_1$-$InS_2I$ | 1.11 | 0.13 | × | | |
| $P2_12_12_1$-$InSe_2F$ | 1.23 | 0.13 | × | | |
| $P2_1$-BaSF | 2.00 | 0.13 | √ | $C2/m$ | 0.07 |
| $P2_1$-$GeBr_2O$ | 2.44 | 0.13 | √ | *Immm* | -1.65 |
| $P2_1$-$OsS_3I_2$ | m | 0.13 | × | | |
| $P2_1$-$SbP_2Cl$ | 0.91 | 0.13 | × | | |
| $P222_1$-$InSe_3Cl_2$ | 0.97 | 0.13 | × | | |
| $P3_1$-$NbSeBr_2$ | 0.17 | 0.13 | × | | |
| $P3_2$-CaPI | 0.89 | 0.13 | × | | |
| $P4_322$_SnSeBr | 0.17 | 0.13 | × | | |
| $P4_3$-GaSCl | 2.77 | 0.13 | √ | *Pnnm* | 0.09 |
| $P2_12_12_1$-$CuSeI_2$ | m | 0.14 | × | | |
| $P2_12_12_1$-$OsSe_2I_2$ | m | 0.14 | × | | |
| $P2_12_12_1$-$TaS_3Cl$ | 0.49 | 0.14 | × | | |
| $P2_13$-$RbBrO_3$ | 4.06 | 0.14 | √ | $R3m$ | 0.10 |
| $P2_1$-$IrS_3Cl_3$ | m | 0.14 | × | | |
| $P2_1$-$TiI_2O$ | 0.70 | 0.14 | √ | *C*2 | -0.02 |
| $P222_1$-$OsSI_3$ | m | 0.14 | × | | |
| $P3_121$-PbSCl | 1.73 | 0.14 | √ | $P3m1$ | -0.27 |
| $P3_121$-$RuI_3Se$ | m | 0.14 | × | | |
| $P3_1$-$SrSe_3I_3$ | m | 0.14 | × | | |
| $P3_1$-TiNF | 1.25 | 0.14 | √ | *Pmmn* | 0.05 |
| $P3_2$-$BaI_2O_3$ | 2.86 | 0.14 | × | | |
| $P3_2$-CuSeI | 1.01 | 0.14 | √ | $P2_1/c$ | 0.09 |
| $P3_2$-$VSI_3$ | m | 0.14 | × | | |
| $P4_12_12$-$BiAs_2F$ | m | 0.14 | × | | |
| $P4_12_12$-$SnAs_2F$ | m | 0.14 | × | | |
| $P4_1$-$BiP_2F$ | m | 0.14 | × | | |
| $P4_32_12$-$BiP_2F$ | m | 0.14 | × | | |
| $P4_322$-InSI | 1.02 | 0.14 | √ | $P2_1/c$ | |
| $P6_1$-$SbBrO_2$ | 0.58 | 0.14 | × | | |
| $P6_5$-$GaS_2F$ | 1.13 | 0.14 | × | | |
| $C222$-$ScSF_2$ | 1.29 | 0.15 | × | | |
| $P2_12_12_1$-$GaS_2I$ | 0.44 | 0.15 | × | | |
| $P2_12_12_1$-InSeI | 0.53 | 0.15 | √ | I41/a | |
| $P2_13$-$ReS_3Cl$ | 0.80 | 0.15 | × | | |
| $P2_1$-$GePI$ | m | 0.15 | × | | |
| $P2_1$-$TlSCl_2$ | 0.66 | 0.15 | × | | |
| $P3_1$-$AlSe_3I_3$ | 1.17 | 0.15 | × | | |
| $P3_1$-$NbS_2Cl_3$ | 0.71 | 0.15 | × | | |
| $P4_12_12$-$SnP_2F$ | m | 0.15 | × | | |
| $P6_2$-$BaSeBr$ | 1.28 | 0.15 | × | | |
| $C2$-$TiSBr_2$ | 1.14 | 0.16 | × | | |
| $P2_1$-$TaSe_2I$ | m | 0.16 | × | | |
| $P3_121$-$OsI_3Se$ | m | 0.16 | √ | $C2/c$ | -0.078 |
| $P3_121$-$RuI_3S$ | m | 0.16 | × | | |
| $P3_1$-$GaSe_3I_3$ | 1.04 | 0.16 | × | | |
| $P4_322$-$IrSe_2I_3$ | m | 0.16 | × | | |
| $P2_12_12_1$-$AuPBr_2$ | 1.38 | 0.17 | × | | |
| $P2_12_12_1$-$TlSe_2Cl$ | 0.25 | 0.17 | × | | |
| $P2_13$-$GaSeBr$ | 0.69 | 0.17 | √ | $P2_1/c$ | 0.16 |
| $P2_13$-$RhOF$ | m | 0.17 | √ | $Pmmn$ | 0.06 |
| $P2_13$-$RuOF$ | 0.01 | 0.17 | √ | $P2_1/c$ | 0.04 |
| $P2_1$-$LaP_2I$ | 0.19 | 0.17 | × | | |
| $P2_1$-$NbBr_2N$ | 1.11 | 0.17 | × | | |
| $P2_1$-$SnP_2Br$ | m | 0.17 | × | | |
| $P3_1$-$HfS_2Br_2$ | 0.87 | 0.17 | √ | $P$-1 | 0.22 |
| $P3_221$-$GeSBr$ | 1.49 | 0.17 | × | | |
| $P3_2$-$CuPBr$ | m | 0.17 | × | | |
| $P4_1$-$InP_2Cl$ | 0.47 | 0.17 | × | | |
| $P4_3$-$NbSe_2F_3$ | 0.90 | 0.17 | × | | |
| $P6_3$-$CuIO_3$ | 1.23 | 0.17 | √ | $R3c$ | -0.06 |
| $P2_12_12_1$-$SbS_2F_2$ | m | 0.18 | × | | |
| $P2_13$-$GaSF$ | 2.92 | 0.18 | √ | $Pnnm$ | 0.02 |
| $P2_13$-$ReSe_3I$ | 0.39 | 0.18 | × | | |
| $P2_13$-$ScSeF$ | 2.44 | 0.18 | √ | $P4/nmm$ | -0.04 |
| $P2_1$-$BiP_2I$ | 0.29 | 0.18 | × | | |
| $P2_1$-$InSBr_2$ | 1.09 | 0.18 | × | | |
| $P3_112$-$AuSCl_2$ | 0.18 | 0.18 | × | | |
| $P3_1$-$SnPBr$ | 0.71 | 0.18 | × | | |
| $P3_2$-$TaSeI_2$ | m | 0.18 | × | | |
| $P4_1$-$AgP_3Br$ | m | 0.18 | × | | |
| $P4_1$-$SbP_2F$ | m | 0.18 | × | | |
| $P4_3$-$PbSe_3I_2$ | 0.24 | 0.18 | × | | |
| $P6_3$-$ZnSe_3Cl_2$ | 0.96 | 0.18 | × | | |
| $P6_4$-$PdSeI$ | m | 0.18 | √ | $P4_2/mmc$ | 0.20 |
| $R3$-$GeP_2Br$ | 1.07 | 0.18 | × | | |
| $C222_1$-$IrSeI_3$ | m | 0.19 | × | | |
| $P2_13$-$YSeBr$ | 2.33 | 0.19 | √ | $F$-43$m$ | 0.12 |
| $P2_1$-LaPF | 0.22 | 0.19 | × | | |
| $P3_2$-$RuSe_2I$ | 0.14 | 0.19 | × | | |
| $P3$-$RuS_3Cl$ | m | 0.19 | × | | |
| $P4_1$-$AgSe_2Cl_3$ | 0.89 | 0.19 | × | | |
| $P4_322$-InSeCl | 0.79 | 0.19 | √ | $P2_1/c$ | 0.19 |
| $P4_3$-GaSeI | 0.60 | 0.19 | √ | $P2_1/c$ | 0.16 |
| $P6_522$-$CoSe_2Br$ | 0.15 | 0.19 | √ | $P2_1/c$ | -0.11 |

**Table S2**. The second-harmonic generation (SHG) coefficients $\chi^{(2)}_{\alpha\beta\gamma}$ (pm/V) of semiconductors in $M_xX_yHal_z$ system with $E_{hull} \leq 20$ meV/atom.

| $\chi^{(2)}_{\alpha\beta\gamma}$ | 111 | 112 | 113 | 123 | 211 | 213 | 233 | 311 | 312 | 323 | 333 |
|---|---|---|---|---|---|---|---|---|---|---|---|
| $P3_1$-$CuBrN_2$ | 2.76 | 32.5 | -34.94 | 71.58 | | | | 10.54 | | | 66.48 |
| $P3_1$-$CuN_2Cl$ | -2.46 | -15.14 | -6.7 | -1.02 | | | | -11.11 | | | 51.78 |
| $P6_3$-$CdSe_3Br_2$ | | | 107.22 | -29.27 | | | | -73.42 | | | -46.43 |
| $P2_13$-$AgBrO_3$ | | | | -3.93 | | | | | | | |
| $P2_13$-$KBrO_3$ | | | | -2.89 | | | | | | | |
| $P2_13$-$NaBrO_3$ | | | | -3.9 | | | | | | | |
| $P2_13$-$NaIO_3$ | | | | -3.52 | | | | | | | |
| $P2_12_12_1$-$AlSe_2Br$ | | | | 7.51 | | 9.82 | | | 0.34 | | |
| $P2_12_12_1$-$AlSe_2Cl$ | | | | 7.1 | | 10.17 | | | 4.38 | | |
| $P2_12_12_1$-$BeS_3Cl_2$ | | | | 17.49 | | -22.86 | | | -1.83 | | |
| $P2_12_12_1$-$CaS_3Cl_2$ | | | | 5.83 | | -19.19 | | | 23.28 | | |
| $P2_12_12_1$-$HfSe_2Br_2$ | | | | -3.02 | | 9.63 | | | -6.1 | | |
| $P2_12_12_1$-$ZrS_2Br_2$ | | | | 5.06 | | 12.2 | | | -15.6 | | |
| $P2_12_12$-$TlS_3Cl$ | | | | 39.01 | | 8.57 | | | -11.6 | | |

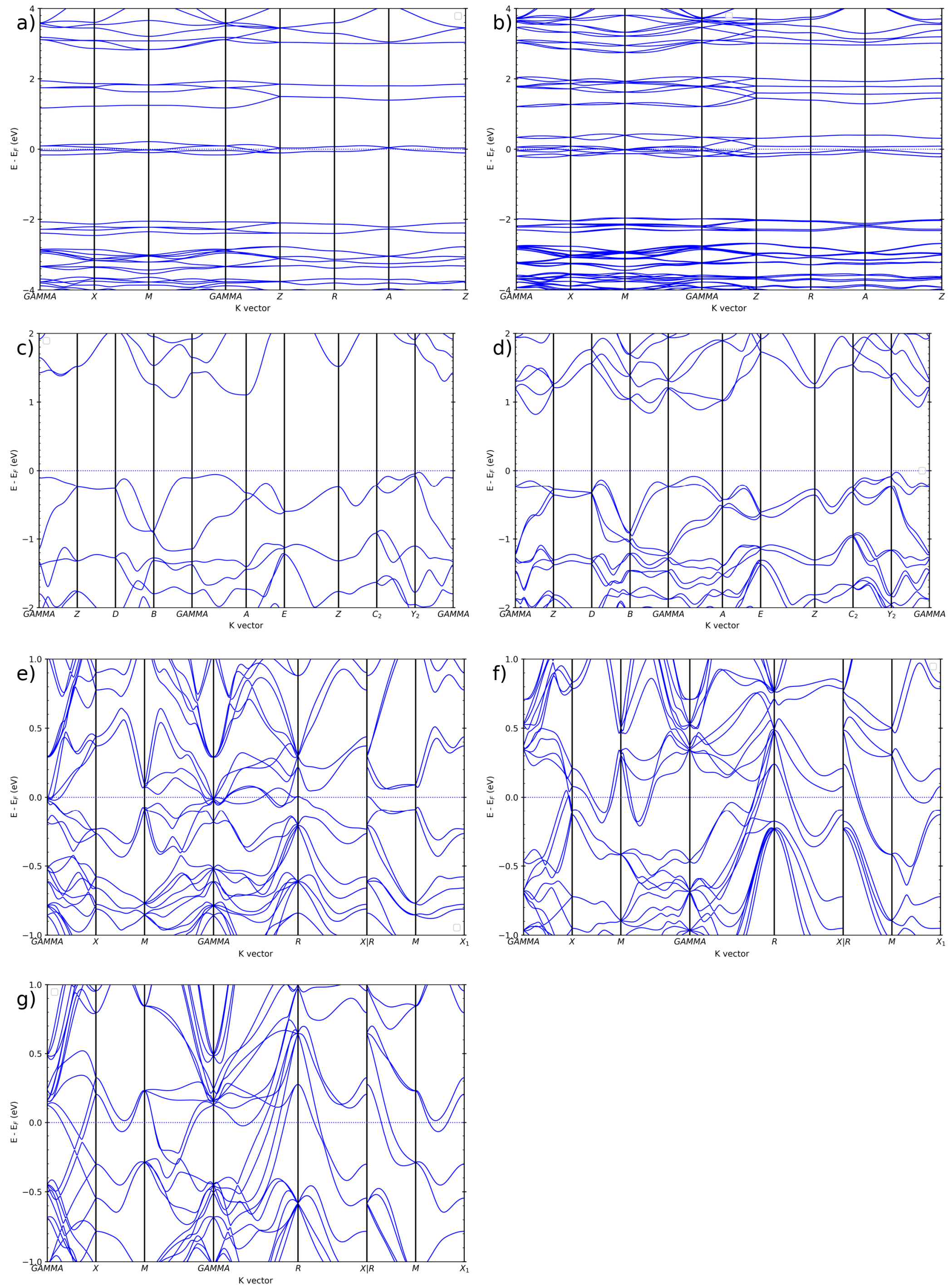

**Fig. S5 Band structures of representative compounds.** The band structure of $P4_1$-$ReNF_3$ without SOC a) and with SOC b). The band structure of $P2_1$-$BiP_2Br$ without SOC c) and with SOC d). The band structure with SOC of e) $P2_13$-$BiBPt_3$, f) $P2_13$-$GeBPt_3$ and g) $P4_332$-$Cd_2BPt_3$.

**Table S3**. The $E_{hull}$ and predicted superconducting transition temperatures (*T*c) and Vickers hardness (*H*v) of Predicted Compound in $M_xM'_yB_z$ system. NAN means that the compound are mechanically unstable based on Ref.[2]

| | $E_{hull}$ (eV/atom) | Target compound exists in OQMD | Space Group In OQMD | δE, $E_{this_work}$-$E_{OQMD}$ (eV/atom) | *T*c (K) | *H*v(GPa) |
|---|---|---|---|---|---|---|
| $I2_12_12_1$-$Na_2Au_2B$ | 0.194 | × | | | 0 | 1.613 |
| $P2_1$-$Li_2InB_2$ | 0.197 | × | | | 3.1 | 2.022 |
| $P2_1$-$Li_3CdB_2$ | 0.184 | × | | | 0.3 | 2.984 |
| $P3_1$-$Li_3CuB_3$ | 0.098 | × | | | 0 | NAN |
| $P4_1$-$Li_3Al_2B$ | 0.183 | × | | | 9.68 | 1.442 |
| $P4_3$-$Cs_2Au_3B$ | 0.153 | × | | | 0 | NAN |
| $P4_3$-$Li_3Zn_2B$ | 0.170 | × | | | 0.913 | 2.647 |
| $P4_132$-$Li_2Au_3B$ | 0.169 | √ | $P4_332$ | 0 | 0 | 0.515 |
| $P4_332$-$Li_2Cu_3B$ | 0.196 | × | | | 0 | 4.618 |
| $P2_12_12_1$-$Be_3CuB$ | 0.187 | × | | | 0 | 40.263 |
| $P2_13$-$Be_3BSb$ | 0.166 | × | | | 0 | 19.620 |
| $P2_13$-$BeAlB$ | 0.0 | × | | | 0 | 28.681 |
| $P3_1$-$BeAl_3B$ | 0.145 | × | | | 0 | 6.595 |
| $P3_2$-$MgAl_3B$ | 0.177 | × | | | 3.98 | 8.333 |
| $P4_2cm$-$Be_3AlB_2$ | 0.061 | × | | | 1.31 | 24.460 |
| $P4_3$-$Be_3AlB$ | 0.143 | × | | | 15.93 | 21.623 |
| $P4_3$-$BeAl_2B$ | 0.175 | √ | *Fm*-3*m* | -0.463 | 1.12 | 9.494 |
| $P4_132$-$Be_3Cu_2B$ | 0.161 | × | | | 0.3 | 21.187 |
| $P2_1$-$La_3SbB_2$ | 0.172 | × | | | 0 | 0.882 |
| $P3_2$-$Y_3Cd_2B_3$ | 0.158 | × | | | 3.31 | 1.640 |
| $P4_132$-$Nb_3Au_2B$ | 0.159 | × | | | 0 | 5.946 |
| $P4_132$-$Nb_3In_2B$ | 0.172 | × | | | 3.396 | 9.634 |
| $P4_332$-$Ta_3Al_2B$ | 0.088 | × | | | 0 | 13.441 |
| $P2_1$-$Al_3ReB$ | 0.178 | × | | | 0 | NAN |
| $P2_1$-$Zn_3Ru_3B_2$ | 0.173 | × | | | 0.515 | 3.589 |
| $C2$-$Ir_3Pb_2B$ | 0.129 | × | | | 2.55 | 3.380 |
| $P2_1$-$Ag_2Rh_3B_2$ | 0.128 | × | | | 0.1 | 3.506 |
| $P2_1$-$Cd_3Rh_3B$ | 0.023 | × | | | 0 | 3.932 |
| $P2_1$-$Cu_2Rh_3B_2$ | 0.199 | × | | | 0.2 | 0.267 |
| $P2_1$-$Tl_3Pt_2B$ | 0.099 | × | | | 5.76 | 0.174 |
| $P2_1$-$TlRh_2B$ | 0.198 | × | | | 1.39 | 0.531 |
| $P2_12_12_1$-$CuPt_3B_2$ | 0.137 | × | | | 1 | 2.028 |
| $P2_12_12_1$-$Ge_2Pt_3B$ | 0.044 | × | | | 0.2 | 3.723 |
| $P2_12_12_1$-$InPt_2B$ | 0.072 | √ | *Fm*-3*m* | -0.766 | 0.4 | 1.145 |
| $P2_12_12_1$-$TlPt_3B$ | 0.009 | √ | *P*6/*mmm* | -0.005 | 0 | 0.673 |
| $P2_12_12_1$-$Zn_2Rh_2B$ | 0.196 | × | | | 0 | 0.791 |
| $P2_12_12$- | 0.139 | × | | | 0.8 | 0.905 |
| $Rh_3Au_3B_2$ | | | | | | |
| $P2_13$-$AlPt_3B$ | 0.055 | √ | *R*-3 | 0.094 | 0.5 | 1.845 |
| $P2_13$-$Pt_3BiB$ | 0.008 | √ | *C*2/*m* | 0.017 | 0 | 0.736 |
| $P2_13$-$Pd_3SbB$ | 0.043 | × | | | 0.5 | 1.440 |
| $P2_13$-$Pt_3GeB$ | 0.011 | × | | | 0 | 0.256 |
| $P2_13$-$Zn_2Rh_3B$ | 0.161 | × | | | 0 | 0.537 |
| $P3_1$-$AlPt_3B_2$ | 0.151 | × | | | 0 | 1.742 |
| $P3_1$-$Cd_2Rh_2B$ | 0.114 | √ | *P*4/*mbm* | 0.002 | 2.76 | 2.057 |
| $P3_2$-$Pt_3Pb_2B$ | 0.108 | × | | | 0 | NAN |
| $P3_2$-$TlPt_2B$ | 0.058 | √ | *Fm*-3*m* | -0.928 | 0 | 1.353 |
| $P4_1$-$Pt_2PbB_2$ | 0.132 | √ | *I*4/*mmm* | 0.049 | 0 | 2.129 |
| $P4_2$-$Pt_3Pb_2B_2$ | 0.080 | × | | | 0 | 1.541 |
| $P4_3$-$Pd_2Au_2B$ | 0.075 | × | | | 0 | 1.188 |
| $P4_3$-$Pt_3Au_2B$ | 0.073 | × | | | 0.1 | 0.476 |
| $P4_3$-$Cd_3Pt_2B$ | 0.148 | × | | | 0 | 1.408 |
| $P4_12_12$-$GePt_2B$ | 0.134 | √ | *Fm*-3*m* | -0.597 | 4.7 | 2.031 |
| $P4_12_12$-$Zn_3Rh_2B$ | 0.184 | × | | | 4.6 | 1.952 |
| $P4_132$-$Zn_2Pd_3B$ | 0.107 | × | | | 0.1 | 2.105 |
| $P4_332$-$Cd_2Pd_3B$ | 0.113 | × | | | 0 | 2.080 |
| $P4_332$-$Cd_2Pt_3B$ | 0.042 | × | | | 0 | 2.062 |
| $P6_1$-$AgPt_2B$ | 0.043 | √ | *Fm*-3*m* | -0.730 | 2.6 | 1.073 |
| $P6_5$-$AgPtB$ | 0.168 | √ | *F*-43*m* | -0.245 | 0 | 1.255 |
| $P6_5$-$BiPd_2B$ | 0.122 | √ | *Fm*-3*m* | -0.734 | 0.7 | 1.280 |
| $P6_5$-$Pd_2PbB$ | 0.149 | √ | *Fm*-3*m* | -0.719 | 0.2 | 1.850 |
| $P6_5$-$SbPd_2B$ | 0.146 | √ | *Fm*-3*m* | -0.622 | 0.3 | 1.913 |

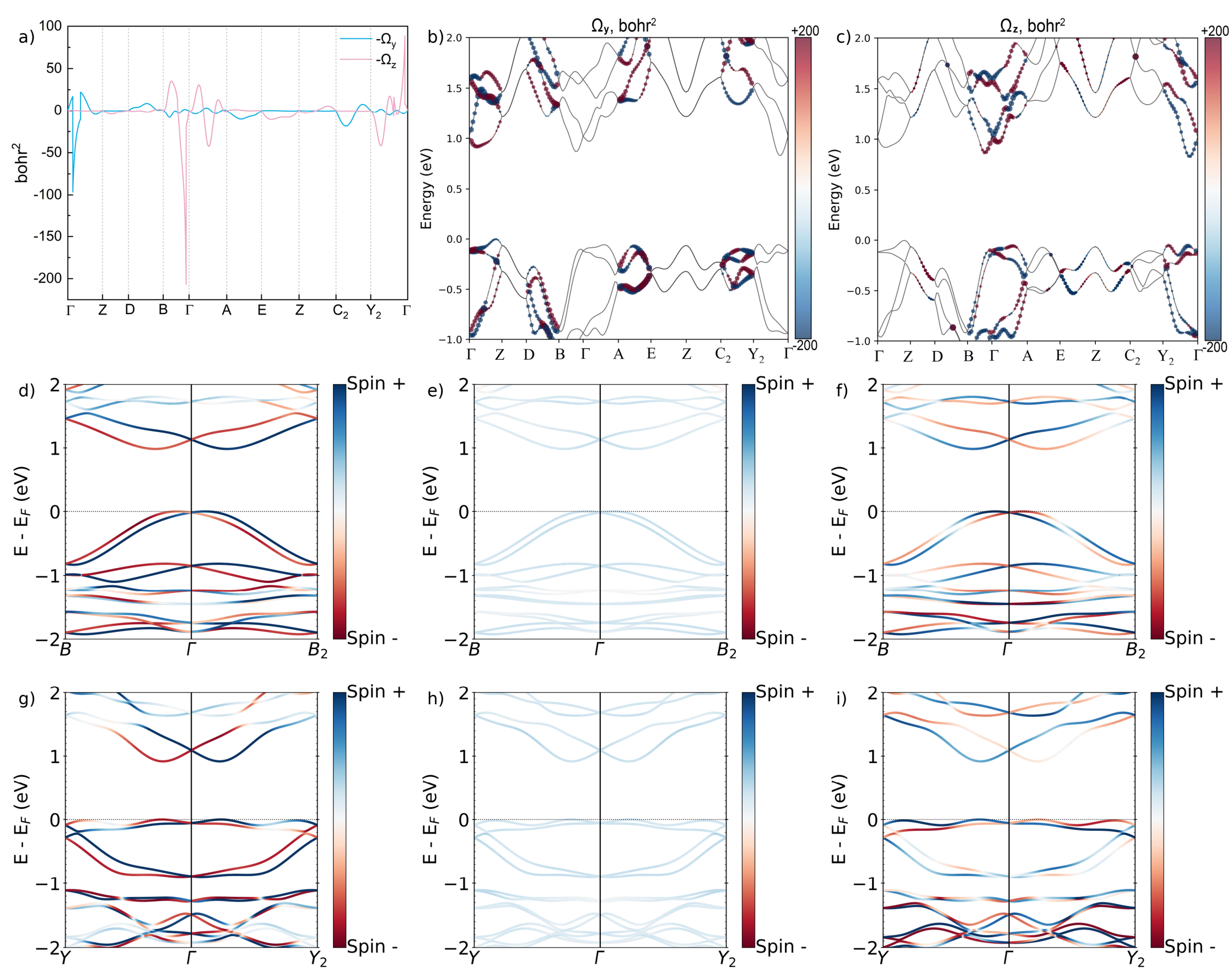

**Fig. S6 Electronic structure of $P2_1$-$BiAs_2Cl$.** a) The integrated Berry curvature of $\Omega_y$ and $\Omega_z$, the Fermi level was tuned as shown in Fig. S2. c. b-c) The band resolved Berry curvature of $\Omega_y$ and $\Omega_z$. d-i) The spin projected band structure of $P2_1$-$BiAs_2Cl$ around Γ point.

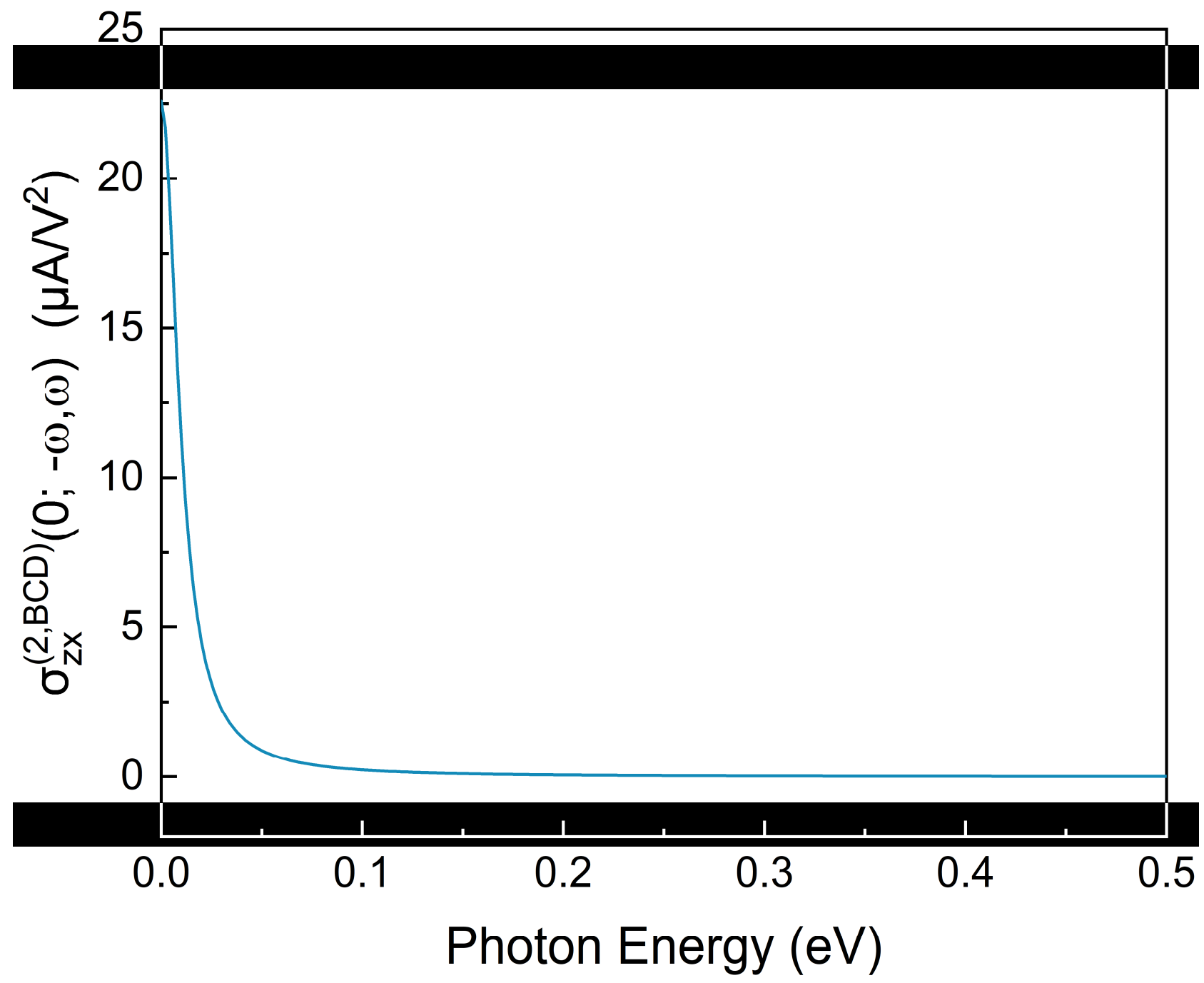

**Fig. S7** The calculated Nonlinear Hall current from BCD contribution fixed at fermi level as shown in Fig S2.c.

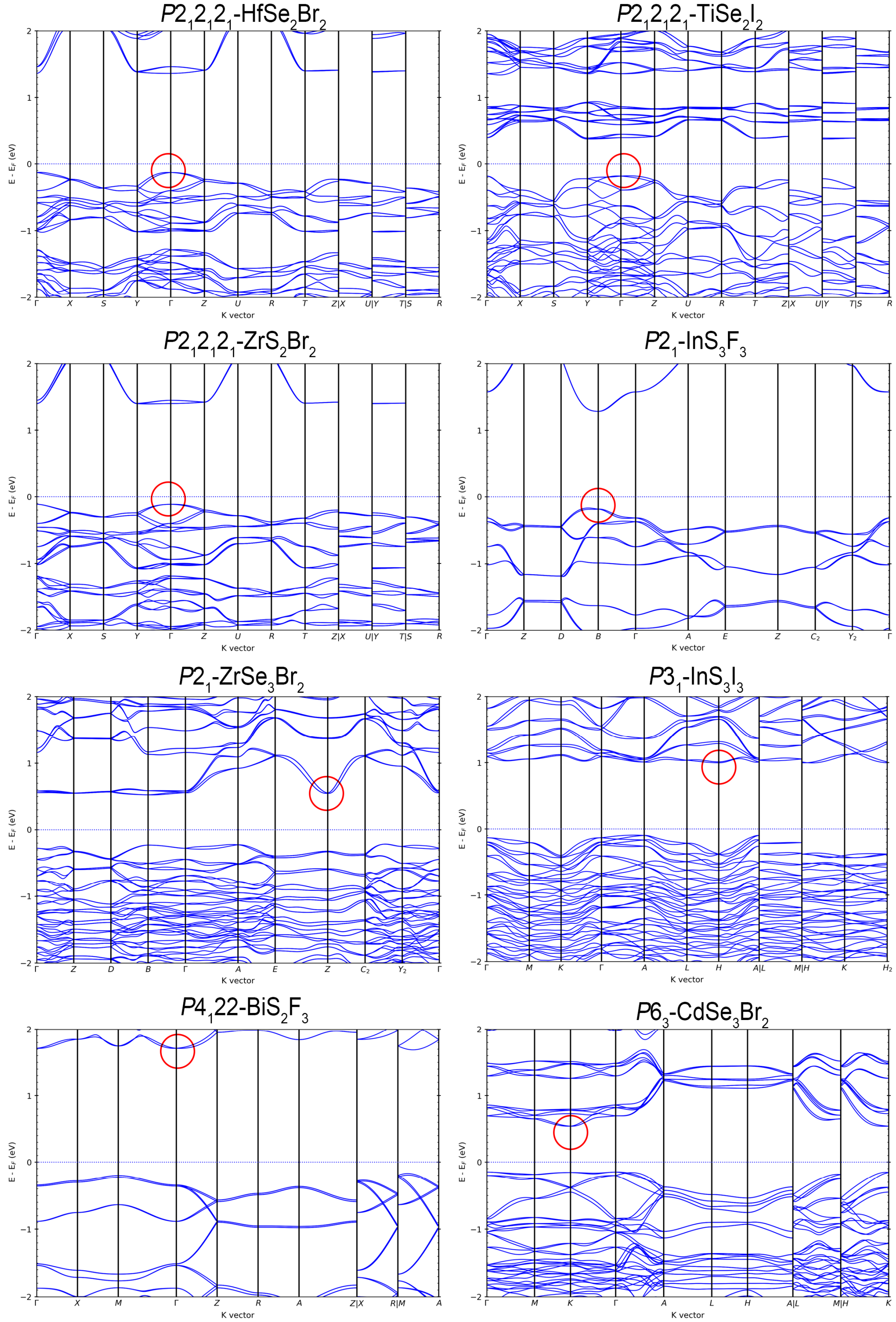

**Fig. S8** Band structures calculated with SOC of representative compounds that have Weyl points at the band edge.

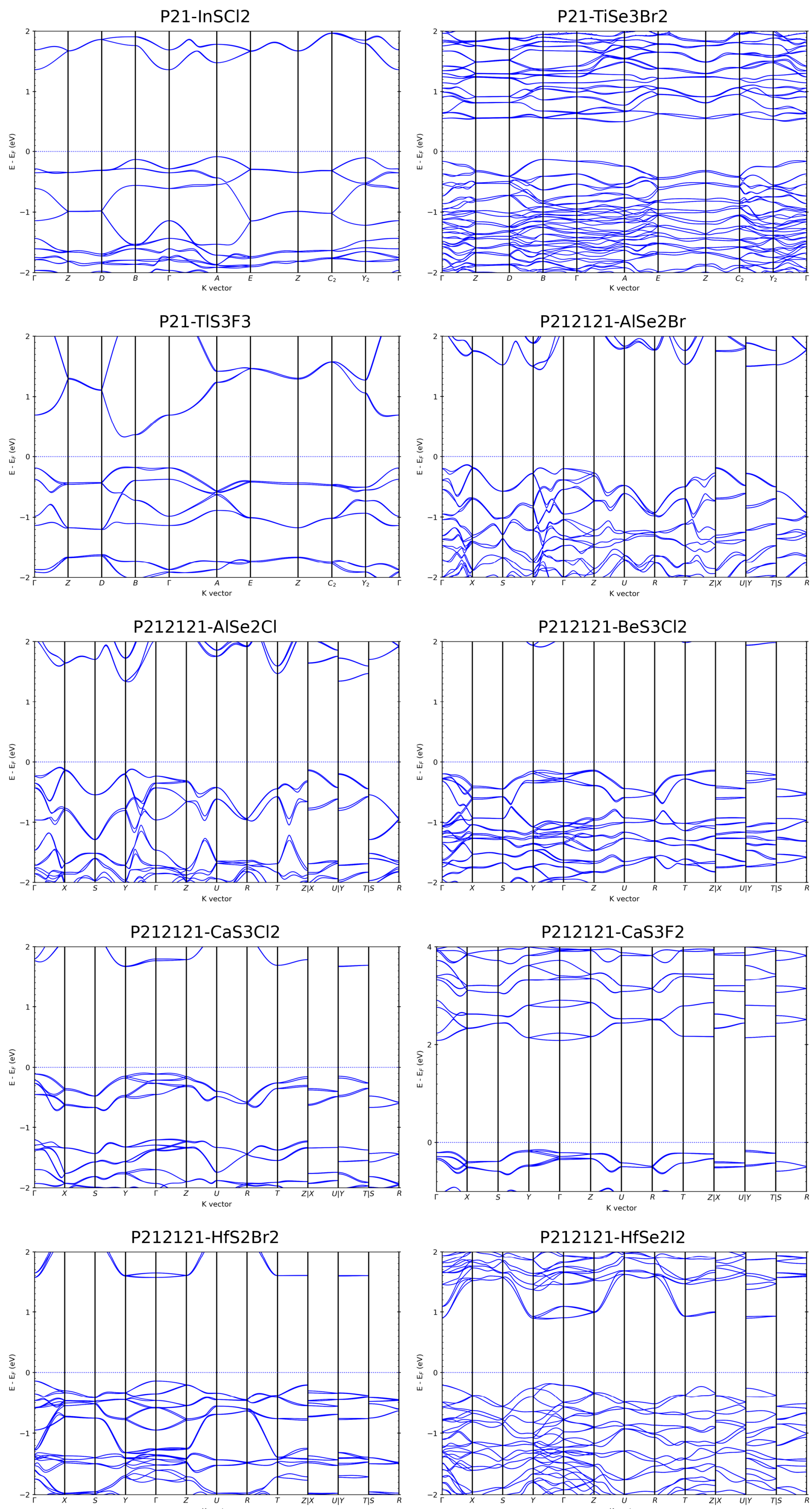

**Fig. S9** Band structures calculated with SOC of semiconductors with $E_{hull} \leq 0.05$ eV/atom.

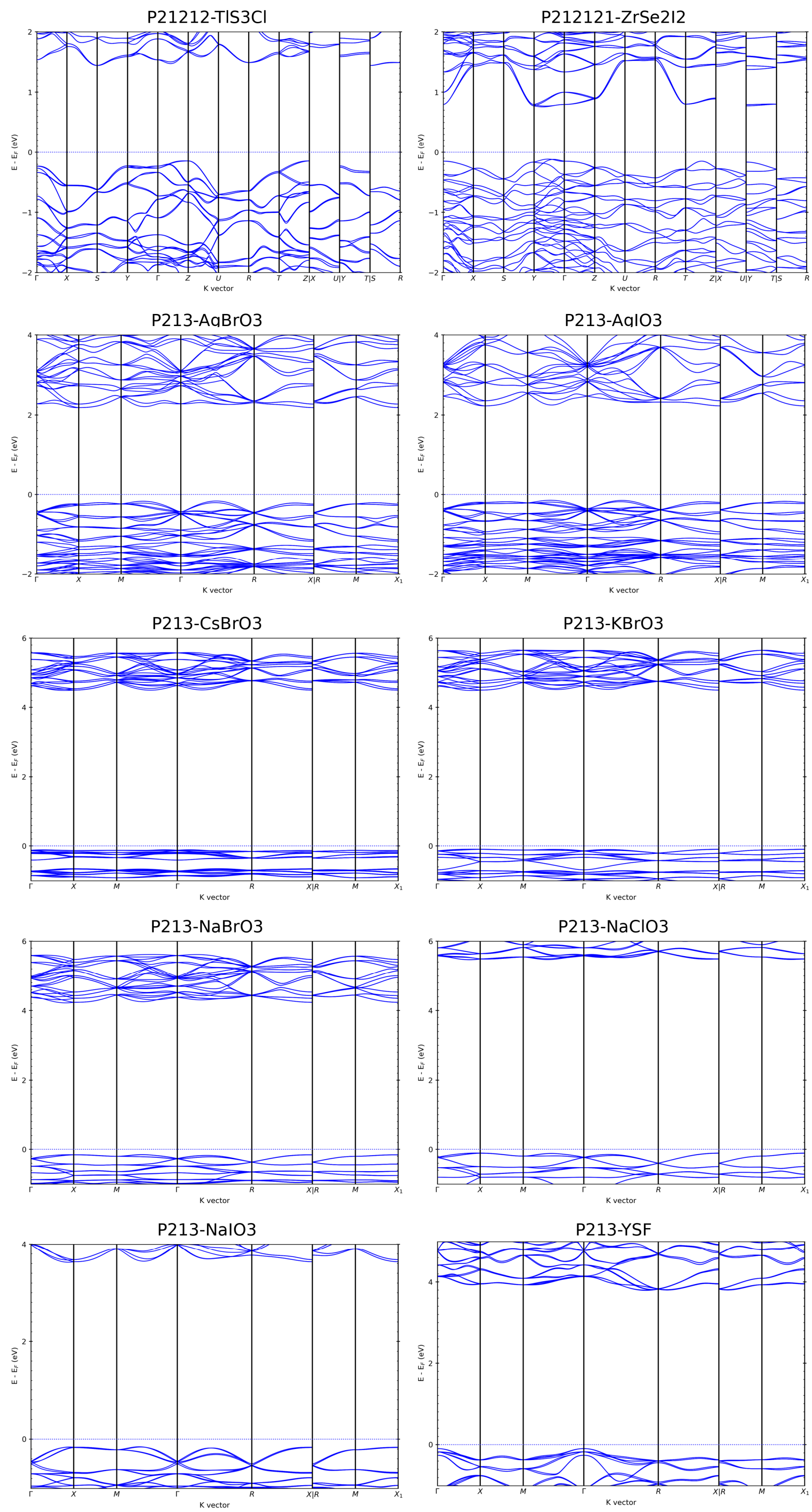

**Fig. S10** Band structures calculated with SOC of semiconductors with $E_{hull} \leq 0.05$ eV/atom.

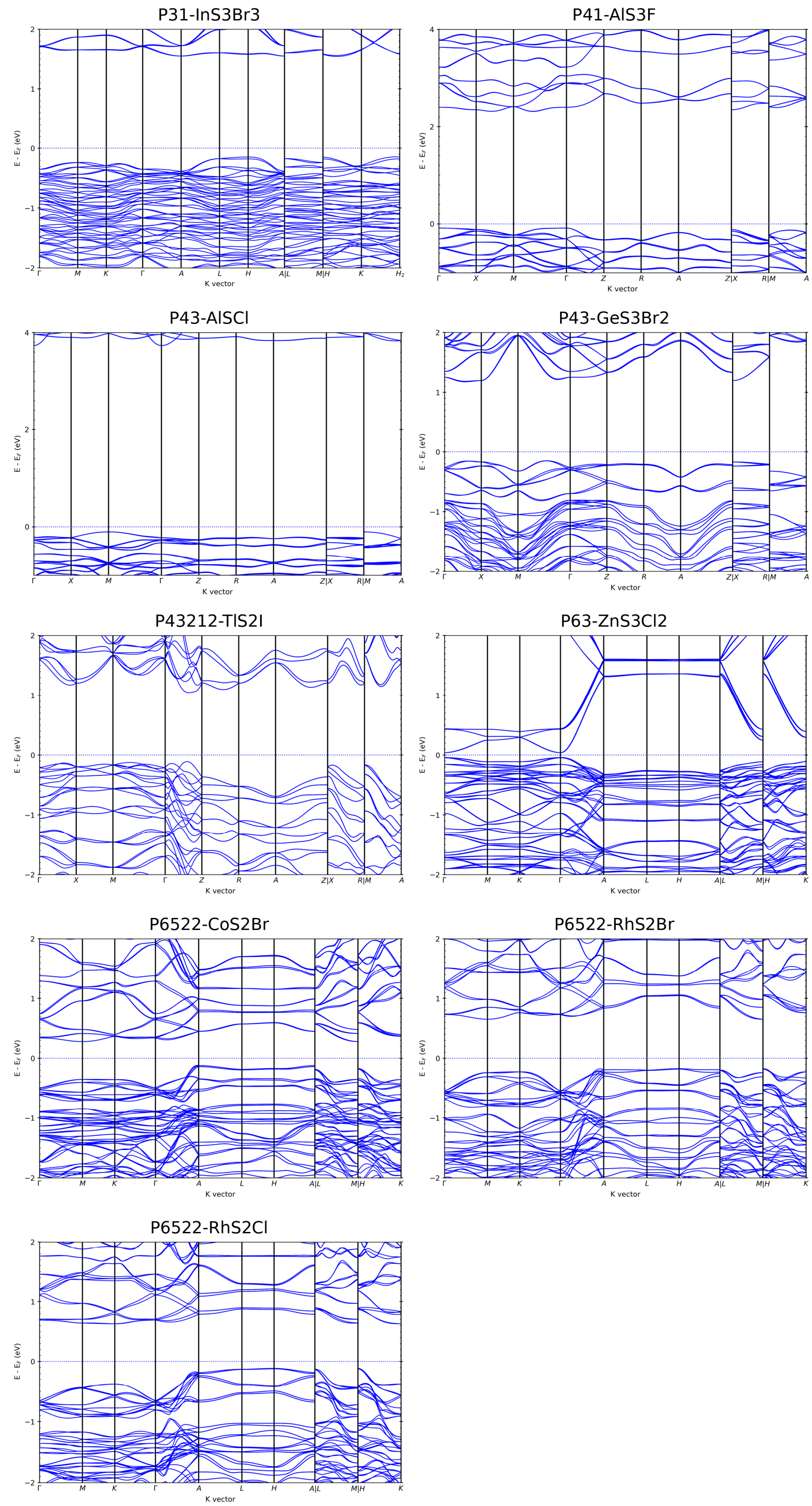

**Fig. S11** Band structures calculated with SOC of semiconductors with $E_{hull} \leq 0.05$ eV/atom.

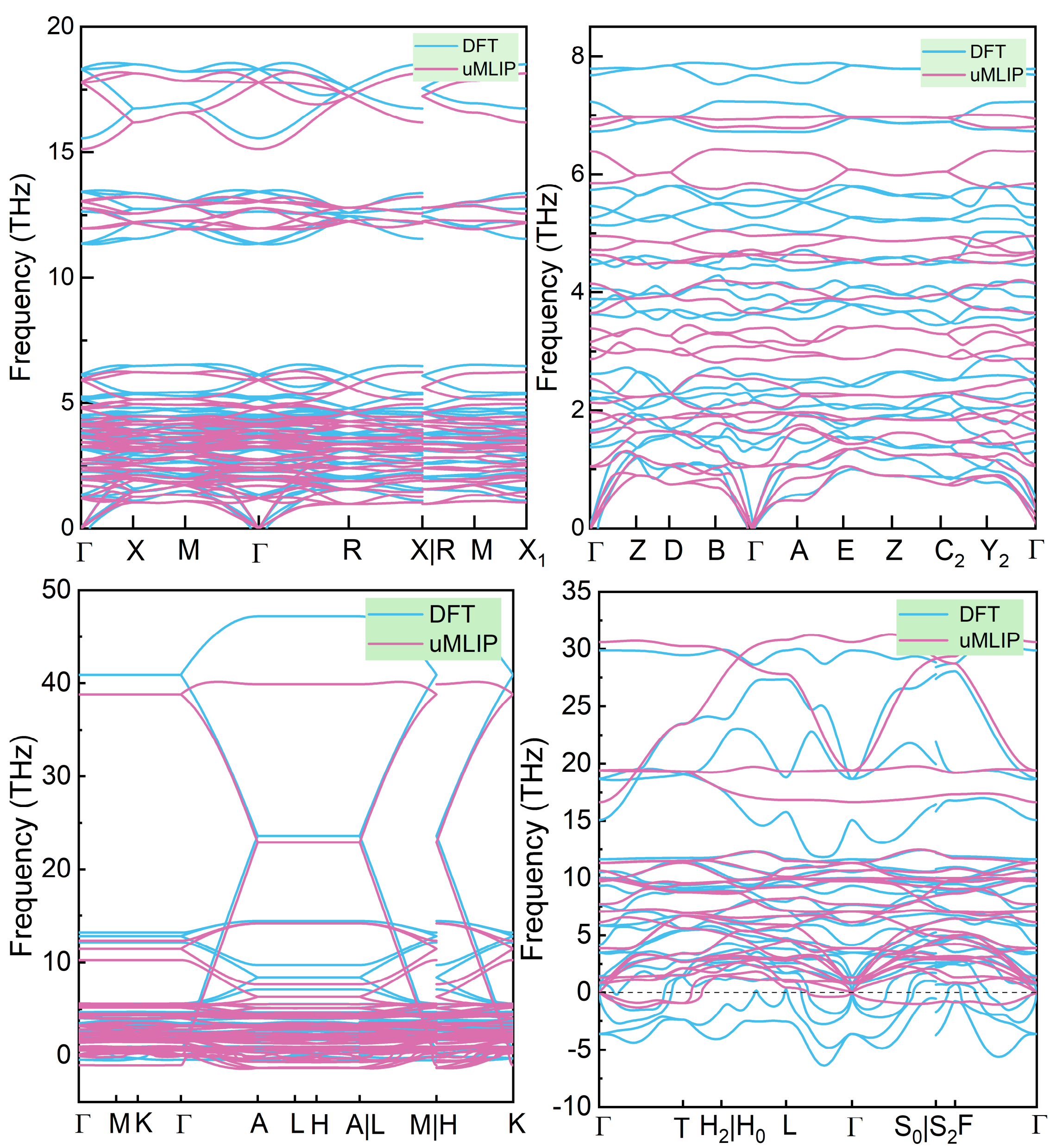

**Fig. S12 Phonon spectra calculated with DFT and uMLIP of** a) $P2_13$-$Pd_3SbB$, b) $P2_1$-$BiAs_2Cl$, c) $P6_2$-$Na_3Tl_2B$ and d) $R32$-$NbClO_2$.

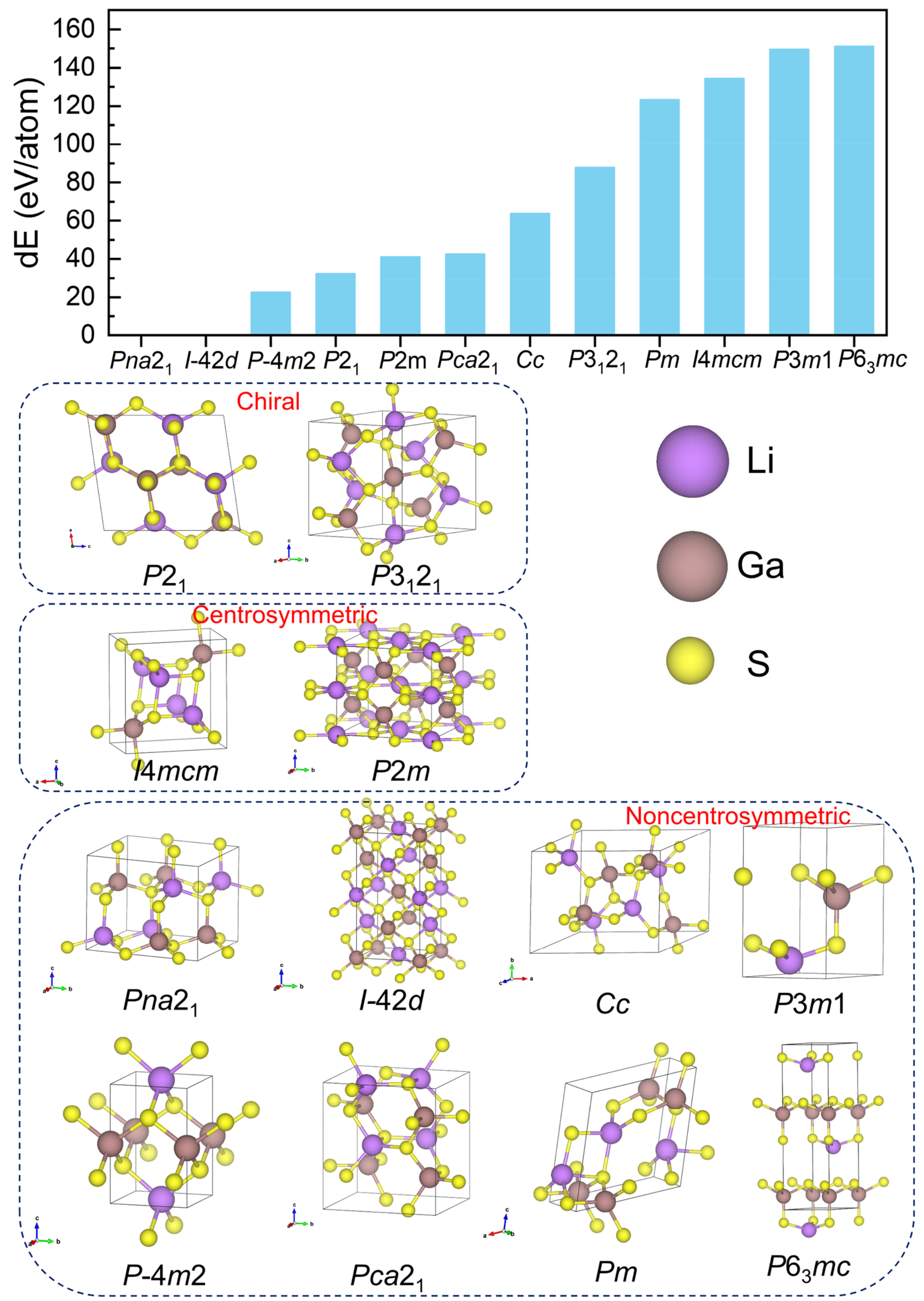

**Fig. S13** A benchmark on $LiGaS_2$ identified several additional low-energy polymorphs that are dynamically stable at the uMLIP level beyond the reported $Pna2_1$ and $I$-42$d$ NLO phase.

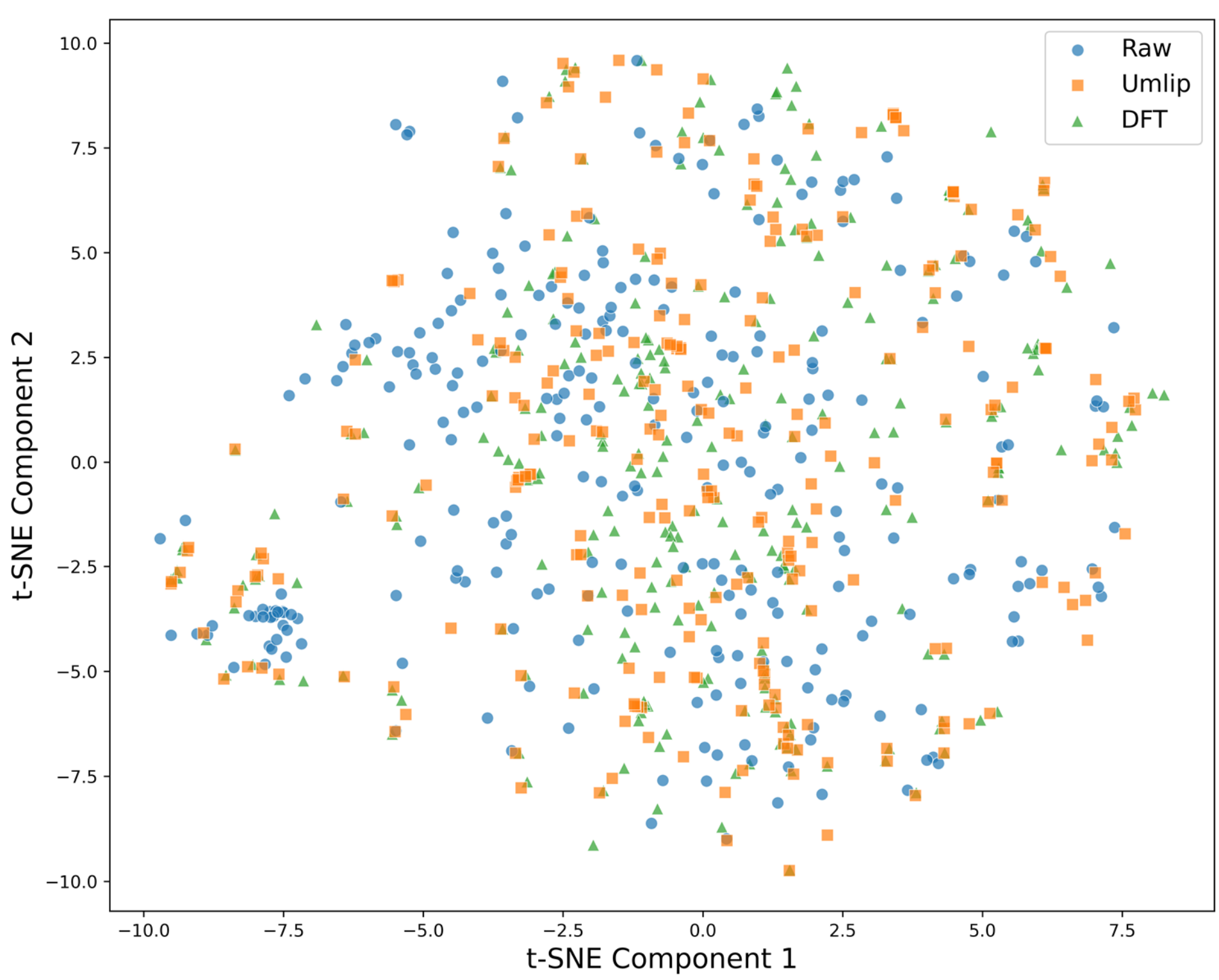

**Fig. S14** The tsne figure with structures after uMLIP (depicted as orange squares) or DFT (represented by green triangles) optimization alongside their corresponding initially randomly generated raw structures (the blue circles).

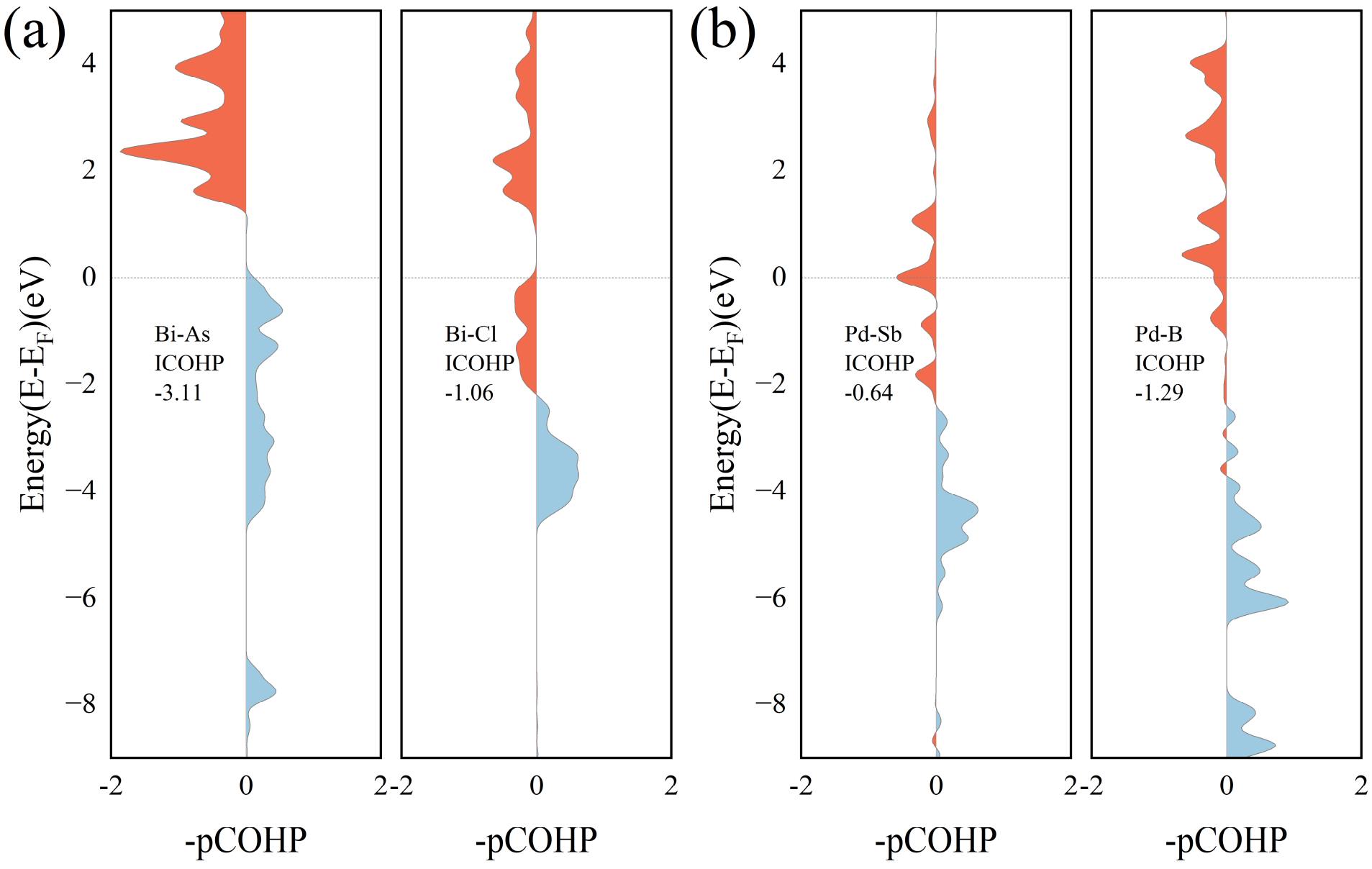

**Fig. S15** Projected Crystal Orbital Hamilton Population (pCOHP) curves for the (a) Bi-As, Bi-Cl interactions in $P2_1$-$BiAs_2Cl$ and (b) Pd-Sb and Pd-B in $P2_13$-$Pd_3SbB$, calculated using LOBSTER based on VASP/PBE wavefunctions.